\documentclass[9pt,twocolumn,twoside]{pnas-new}
\usepackage{xr}
\makeatletter
\newcommand*{\addFileDependency}[1]{
  \typeout{(#1)}
  \@addtofilelist{#1}
  \IfFileExists{#1}{}{\typeout{No file #1.}}
}
\makeatother

\newcommand*{\myexternaldocument}[1]{
    \externaldocument{#1}
    \addFileDependency{#1.tex}
    \addFileDependency{#1.aux}
}
\myexternaldocument{PNAS-NFL-SI}

\setboolean{displaywatermark}{false}
\templatetype{pnasresearcharticle} % Choose template 
\usepackage{amsfonts}
\usepackage{upgreek}
\usepackage{slashed}
\usepackage{latexsym}
\usepackage[export]{adjustbox}
\usepackage[colorlinks=true]{hyperref}  
\usepackage[toc,page]{appendix}
\usepackage{dsfont}
\usepackage{pdfpages}
\usepackage{graphicx}
\usepackage{subfig}
\def\bea{\begin{eqnarray}}
\def\eea{\end{eqnarray}}
\def\beq{\begin{equation}}
\def\eeq{\end{equation}}

\newcommand{\iw}{i\omega}
\newcommand{\om}{\omega}

\newcommand{\rs}{\rho_{s}} 

\newcommand{\bs}{\mathfrak{b}}
\newcommand{\fh}{\mathfrak{f}}
\title{Critical metallic phase in the overdoped random $t$-$J$ model}

\author[a]{Maine Christos}
\author[a]{Darshan G. Joshi} 
\author[a,b]{Subir Sachdev}
\author[a]{Maria Tikhanovskaya}

\affil[a]{Department of Physics, Harvard University, Cambridge MA-02138, USA}
\affil[b]{School of Natural Sciences, Institute for Advanced Study, Princeton, NJ-08540, USA}
\leadauthor{Christos} 

\significancestatement{The strange metal state of the cuprate superconductors is the parent state from which many novel quantum phases emerge. Experiments have shown that the strange metal extends over a wider range of electron densities than previously believed. We show that a dynamic mean-field theory with random exchange interactions, in which the electron fractionalizes into excitations carrying its spin and charge, displays an extended strange metal regime. Several physical properties of this strange metal are in accord with experimental observations, including a linear-in-temperature contribution to the resistivity. At lower carrier density, our theory displays an instability to a spin glass, which is also observed. Our mean-field theory points a route towards theories of quantum materials based upon more realistic microscopic models.}

\authorcontributions{Author contributions: All authors conceived the project, performed the theoretical analysis and wrote the paper. MC, DGJ, and MT performed the numerical analyses.}
\authordeclaration{The authors declare no competing interest.}
\equalauthors{\textsuperscript{1}To whom correspondence may be addressed. Email: sachdev@g.harvard.edu}
\keywords{Strange metal $|$ Spin glass $|$ Random $t$-$J$ model} 

\begin{abstract}
We investigate a model of electrons with random and all-to-all hopping and spin exchange interactions, with a constraint of no double occupancy. The model is studied in a Sachdev-Ye-Kitaev-like large-$M$ limit with SU($M$) spin symmetry. The saddle point equations of this model are similar to appoximate dynamic mean field equations of realistic, non-random, $t$-$J$ models.
We use numerical studies on both real and imaginary frequency axes, along with asymptotic analyses, to establish the existence of a critical non-Fermi-liquid metallic ground state at large doping, with the spin correlation exponent varying with doping. This critical solution possesses a time-reparametrization symmetry, akin to SYK models, which contributes a linear-in-temperature resistivity over the full range of doping where the solution is present. It is therefore an attractive mean-field description of the overdoped region of cuprates, where experiments have observed a linear-$T$ resistivity in a broad region. The critical metal also displays a strong particle-hole asymmetry, which is relevant to Seebeck coefficient measurements. We show that the critical metal has an instability to a low-doping spin-glass phase, and compute a critical doping value. We also describe the properties of this metallic spin-glass phase.
\end{abstract}

\dates{This manuscript was compiled on \today}
\doi{\url{www.pnas.org/cgi/doi/10.1073/pnas.XXXXXXXXXX}}

\begin{document}

\maketitle
\thispagestyle{firststyle}
\ifthenelse{\boolean{shortarticle}}{\ifthenelse{\boolean{singlecolumn}}{\abscontentformatted}{\abscontent}}{}

% If your first paragraph (i.e. with the \dropcap) contains a list environment (quote, quotation, theorem, definition, enumerate, itemize...), the line after the list may have some extra indentation. If this is the case, add \parshape=0 to the end of the list environment.
\dropcap{R}ecent experimental works have highlighted certain fundamental properties of cuprate superconductors and their complex and rich phase diagrams. One of the key aspects is a transformation in the normal state near an optimal doping $p=p_{c}$ \cite{Proust_review, Michon2019, badoux2016, arpes_chen2019} indicated most recently by thermal-Hall transport measurements \cite{Proust_review, badoux2016} and by photoemission experiments \cite{arpes_chen2019}. While much attention has focused on the anomalous properties of the underdoped regime ($p<p_c$), it is often assumed that the overdoped regime ($p>p_{c}$) is a conventional Fermi liquid, and thus the latter has not attracted as much attention. However, careful experimental studies have reported significant strange-metal anomalies in transport properties also on the overdoped side \cite{Cooper09,Greene20,Hussey21}. These observations indicate the presence of a non-Fermi-liquid metal in an extended doping region above optimal doping. On the underdoped side, recent experiments have established that there is a spin glass phase \cite{Rainford2004, Frachet2020} in the La-based compounds.

Concordant with these observations, we present here a theoretical model with an extended non-Fermi-liquid phase at large dopings, and a spin-glass phase at low doping. 

Models with random interactions on a fully connected lattice in the Sachdev-Ye-Kitaev class \cite{SY93, kitaev2015talk} yield a systematic route to studying non-Fermi liquids: the exact solutions of such models serve as dynamic mean-field theories of more realistic microscopic models \cite{Haule1,Haule2,syk_rmp}. 
In this paper we report numerical and analytic solutions of a 
$t$-$J$ model (Eq. \ref{eq:Ham}) with random and all-to-all hopping and exchange interactions across a wide range of doping. The model has a global SU($M$) spin rotation symmetry, and we study a particular SYK-like large $M$ limit with fermionic spinons.  A previous study \cite{Joshi:2019csz} found possible critical solutions of non-Fermi liquid metals by an analytic study of the low energy limit of the saddle point equations of this large $M$ limit. From renormalization group arguments it appeared that these possible critical solutions only described a critical point or a small range of intermediate doping, and that a Fermi liquid solution would appear in the overdoped regime. 

%%%%%%%%%%%%%%%%%%%%%%%%%%%%%%%%%%%%%
\begin{figure}[t]
\centering
    \includegraphics[width=0.94\linewidth]{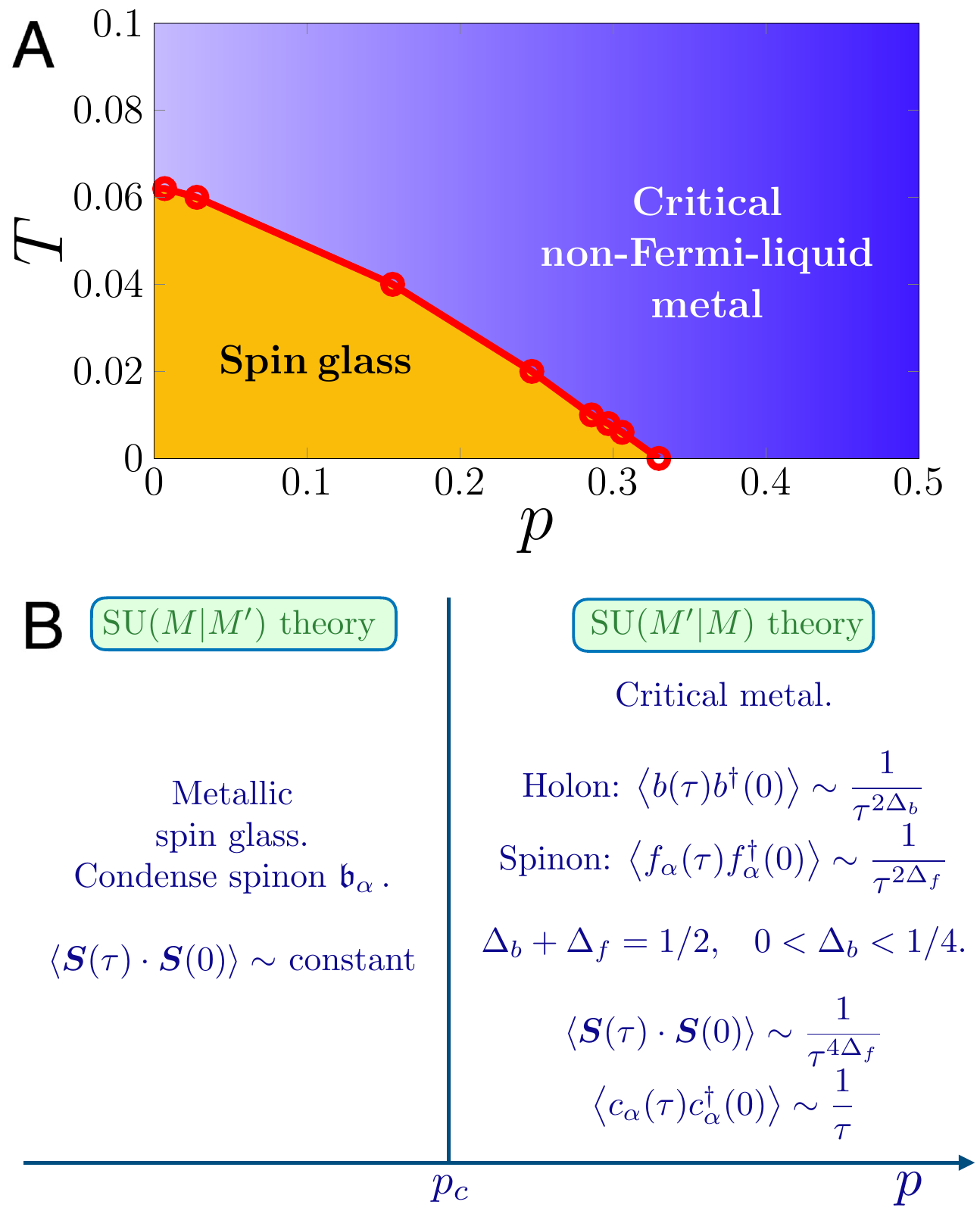}
    \caption{
    (A) Phase diagram as a function of doping $p$ and temperture $T$ obtained from numerically solving the fermionic spinon large $M$ saddle-point equations of the random $t$-$J$ model in Eq. \ref{eq:Ham} for $t=J=1$. The red circles are critical doping value ($p_c(T)$) at a given temperature, obtained by looking at the spin-glass instability in the non-Fermi-liquid solution. The uncertainty for each $p_c$ is within $0.003$. The critical doping at $T=0$ is obtained by directly solving the saddle-point equations on the real-frequency axis. (B) Properties of the $T=0$ states. The low-doping spin-glass phase and large-doping critical-metal phase are separated by a quantum critical point at $p_{c}$. The critical metal has doping-dependent exponents $\Delta_{f,b}$, and a linear-$T$ resistivity.}
    \label{fig:pd}
\end{figure}
%%%%%%%%%%%%%%%%%%%%%%%%%%%%%%%%%%%%%

The present paper will present a full numerical solution of these large $M$ saddle point equations for a wide range of doping. The solutions are obtained on both the real and imaginary frequency axes, with mutually consistent results. We also supplement the numerical results with asymptotic analytic analyses. Our main results are 
that the Fermi liquid is never the solution of the large-$M$ saddle-point equations, and that one of the low energy critical solutions obtained earlier \cite{Joshi:2019csz} extends to an all-energy solution of the large-$M$ equations in the entire overdoped regime. We further show that there is a phase transition to a low-doping spin-glass phase and construct a phase diagram (see Fig. \ref{fig:pd}). 

Notable features of the novel critical non-Fermi liquid phase in Fig.~\ref{fig:pd} are \\
({\it i\/}) spin correlations decay with an exponent which varies continuously with doping (unlike a Fermi liquid),\\
({\it ii\/}) the electron correlators have the $\sim 1/\tau$ decay with imaginary time (as in a Fermi liquid), but with a pronounced particle-hole asymmetry (which is weak in a Fermi liquid), and\\
({\it iii\/}) the mechanism of Ref.~\cite{Guo:2020aog} applies across the entire overdoped critical phase, and the resistivity has a linear-$T$ contribution as $T \rightarrow 0$ at all dopings in this phase (unlike the $T^2$ resistivity in a Fermi liquid).

We also note here that there is a distinct large $M$ limit of the random $t$-$J$ model \cite{PG99} which yields a Fermi liquid ground state at all non-zero doping. We will discuss the relation to this limit in Section~\ref{sec:disc}.

The plan of the paper is as follows. In Section~\ref{sec:model} we introduce the model and discuss its general properties. In Section~\ref{sec:overd} we consider the critical solutions of the model introduced in Eq. \ref{eq:Ham}, and show that there is a critical solution with doping-dependent exponents in the overdoped region. We then solve this model at zero temperature (Section~2.\ref{sec:realw}), as well as finite temperature (Section~2.\ref{sec:imagw}), and show that it has an instability to a spin-glass phase. The results for spectral functions in the spin-glass phase are presented in Section~\ref{sec:underd} using an alternative bosonic spinon large $M$ limit. In both phases we report physical observables such as electron and spin spectral densities at both zero and finite temperatures. We conclude with a discussion and implication of our results in Section~\ref{sec:disc}. Technical details and additional results are presented in the appendices. 

%%%%%%%%%%%%%%%%%%%%%%%%%%%%%%%%%%%%%%%%%%%%%%%%%%%%%%%%%%%%%
\section{Model}
\label{sec:model}

We consider a model of electrons with random and all-to-all hopping and exchange interaction with double occupancy being prohibited. This is the random $t$-$J$ model which considers doping a random Heisenberg magnet \cite{SY93}. It is in the class of SYK models \cite{SY93, kitaev2015talk} and is suitable for studying metallic phases obtained upon doping a Mott insulator. The Hamiltonian is
\begin{equation}
\label{eq:Ham}
H = \frac{1}{\sqrt{N}} \sum_{i \neq j =1}^{N} t_{ij} c_{i\alpha}^{\dagger} c_{j\alpha}  
+ \frac{1}{\sqrt{N}} \sum_{i < j =1}^{N} J_{ij} \vec{S}_{i}\cdot\vec{S}_{j} 
- \mu \sum_{i} c_{i\alpha}^{\dagger} c_{i\alpha} \,,
\end{equation}
with the constraint, $\sum_{\alpha} c_{i\alpha}^{\dagger} c_{i\alpha} \leq 1$, since double occupancy is not allowed. In the above Hamiltonian, $c_{\alpha}$ is the electron annihilation operator with $\alpha = \uparrow \,, \downarrow$, the spin operator is $S_{i}^{a} = c_{i\alpha}^{\dagger} \sigma^{a}_{\alpha\beta} c_{i\beta}/2$, and $\mu$ is the chemical potential.  
The complex hoppings $t_{ij}$ and real exchange interactions $J_{ij}$ are independent random numbers with zero mean and mean-square values $t^{2}$ and $J^{2}$ respectively. 

Note that one could also consider a $t$-$J$ model with non-random nearest-neighbor hopping and nearest-neighbor random exchange interactions on a large dimensional lattice. The on-site dynamical mean-field equations for such a model are the same as those obtained for the model in Eq. \ref{eq:Ham}. It also allows for a definition of transport quantities such as resistivity. 

We will consider the problem at finite hole doping ($p$). This was recently studied both analytically \cite{Joshi:2019csz, Tikhanovskaya:2020zcw} and numerically \cite{Shackleton2021,Dumitrescu2021}. In particular, a deconfined critical point scenario was proposed in Ref. \cite{Joshi:2019csz} to describe a quantum phase transition between a spin-glass phase at low doping with carrier density $p$ and a Fermi-liquid phase at large doping with carrier density $1+p$. The deconfined critical point proposed therein had the property that the local spin susceptibility was marginal as in the SYK models \cite{SY93, kitaev2015talk}. This was based on renormalization group arguments. However, a large-$M$ analysis \cite{Joshi:2019csz} led to two types of critical metallic solutions. One of the critical solutions corresponds to the deconfined critical point (although it is a phase in large-$M$ limit), while the other was believed to be suppressed in favor of a Fermi-liquid phase.

In this work we find that, in fact, the second critical solution is stable, and a Fermi-liquid phase is never achieved within a large-$M$ approach at the saddle-point level. This critical phase has the property that while the exponent of the spin correlation continuously varies with doping, the linear-$T$ resistivity is present over the entire overdoped phase. This makes it an attractive candidate for the overdoped phase of cuprate in the light of recent experiments \cite{Cooper09, Hussey21}. In the underdoped region, below a critical doping $p_{c}$, we find a spin-glass phase. Thus, the critical metal at large doping and spin-glass phase at small doping are separated by a quantum critical point at a finite doping $p_{c}$, as shown in Fig. \ref{fig:pd}.

As a result of the double-occupancy constraint each site has three states, namely empty ($\left|0\right\rangle$) and singly-occupied ($\left|\uparrow\right\rangle$ and $\left|\downarrow\right\rangle$) states. These can be conveniently described using holon and spinon operators. The electron is thus fractionalized into holon and spinons. The critical metallic solutions in the overdoped region has gapless and critical fermionic spinons and bosonic holons (Section~\ref{sec:overd}), while the underdoped spin-glass phase will be described by a fermionic holon and bosonic spinons (Section~\ref{sec:underd}). 

%%%%%%%%%%%%%%%%%%%%%%%%%%%%%%%%%%%%%%%%%%%%%%%%%%%%%%%%%%%%%
\section{Critical non-Fermi-liquid metal phase}
\label{sec:overd}

In this section, we show that our model in Eq. \ref{eq:Ham} admits a critical metallic solution. This phase is described by fermionic spinon ($f_{\alpha}$) and bosonic holon ($b$) operators. The electron and spin operators can be then written in terms of these fractional particles as
\begin{equation}
\label{eq:cs1}
c_{i\alpha} =  b^{\dagger} f_{i\alpha} \,, ~~
S^{a}_{i} = f_{i\alpha}^{\dagger} \frac{\sigma^{a}_{\alpha\beta}}{2} f_{i\beta} \,, 
\end{equation}
with the constraint, $f_{i\alpha}^{\dagger} f_{i\alpha} + b_{i}^{\dagger} b_{i} =1$. This theory realizes a SU$(1|2)$ superalgebra. 
The strategy to solve the model in Eq. \ref{eq:Ham} is to generalize to a larger symmetry. The spin index on the spinon operator ($f_{\alpha}$) is generalized to $\alpha = 1,\ldots,M$ and there is an additional orbital index for holon ($b_{\ell}$) such that $\ell = 1,\ldots, M'$. This theory realizes a SU$(M'|M)$ superalgebra. In this larger symmetry the electron and spin operators take the form,
\begin{equation}
\label{eq:cs1M}
c_{\ell \alpha} =  b^{\dagger}_{\ell} f_{i\alpha} \,, ~~ 
S^{a} = f_{\alpha}^{\dagger} T^{a}_{\alpha\beta} f_{\beta} \,, 
\end{equation}
where the matrices $T^{a}$ obey SU$(M)$ algebra. The constraint now takes the general form,
\begin{equation}
\label{eq:fb_const}
f_{\alpha}^{\dagger} f_{\alpha} + b_{\ell}^{\dagger} b_{\ell} = \kappa M \,. 
\end{equation}
The idea is to take a large-$M', M$ limit such that $k=M'/M$ is finite. This large $M$ limit is taken {\it after} we have performed a disorder average, and taken the large-volume limit. Note that this large $M$ limit is distinct from the large $M$ limit in Ref.~\cite{PG99}, which had $M'=1$; the latter limit leads to a Fermi liquid phase at $T=0$ at all non-zero doping.

A detailed analysis of our large $M$ limit can be found in Refs. \cite{Joshi:2019csz, Tikhanovskaya:2020zcw} and here we simply recall the saddle-point equations derived therein, which are 
%Recall the fermionic spinon theory of the {\it random\/} $t$-$J$ from Ref.~\cite{Tikhanovskaya:2020zcw}:
\begin{align}
  &G_b(i\omega_n) = \frac{1}{i\omega_n+\mu_b-\Sigma_b(i\omega_n)}, \label{Eq:EoM1} \\
  &\Sigma_b(\tau) = -t^2G_f(\tau)G_f(-\tau)G_b(\tau) \label{Eq:EoM2},\\
  &G_f(i\omega_n) = \frac{1}{i\omega_n+\mu_f-\Sigma_f(i\omega_n)} \label{Eq:EoM3},\\
  &\Sigma_f(\tau) = -J^2 G_f(\tau)^2G_f(-\tau)+kt^2G_f(\tau)G_b(\tau)G_b(-\tau) \label{Eq:EoM4},
\end{align}
where $G_{b}$ and $G_{f}$ are the boson and fermion Green's function respectively, while $\Sigma_{b}$ and $\Sigma_{f}$ are the boson and fermion self energies respectively. 
Here $\tau$ is the imaginary time and $\om_{n}$ are the Matsubara frequencies. The chemical potentials $\mu_f$ and $\mu_b$ are 
%determined by the saddle point value of $\lambda$, 
chosen to satisfy
\beq
\left\langle f^\dagger f \right\rangle = \kappa-k p \quad, \quad  \left\langle b^\dagger b \right\rangle = p \,. \label{lutt1}
\eeq
We will restrict our attention to the physical case, $\kappa=1/2$ and $k=1/2$.
In terms of the holon and spinon Green's functions, the electron Green's functions and the spin correlator are,
 \begin{align}
  &   G_{c} (\tau) = -\langle T_{\tau} c_{\alpha}(\tau) c^{\dagger} (0) \rangle = - G_f (\tau) G_b (-\tau) \\
   &  \chi (\tau) = \langle \vec{S}(\tau)\cdot \vec{S}(0) \rangle = - G_f (\tau) G_f (-\tau) \,.
  \label{Eq:EoM5}
 \end{align}
Similar equations have been obtained by a dynamic mean field theory of a non-random $t$-$J$ model on the square lattice using a `non-crossing' approximation, and studied numerically \cite{Haule1,Haule2}. 

To solve the saddle-point equations on the imaginary-frequency axis, it is convenient to define
%For the imaginary time solution, let us define
\beq
\beta \overline{r} = - G_b (i \omega_n = 0) \,.
\eeq
Then we can eliminate $\mu_b$ and obtain the following set of equations to solve
\begin{align}
&G_b(i\omega_n) = \frac{1}{i\omega_n - 1/( \beta \overline{r}) -\Sigma_b(i\omega_n)+ \Sigma_b(i \omega_n=0)} \label{f1} \\
 & \Sigma_b(\tau) = -t^2G_f(\tau)G_f(-\tau)G_b(\tau) \label{f2} \\
  &G_f(i\omega_n) = \frac{1}{i\omega_n+\mu_f-\Sigma_f(i\omega_n)} \label{f3}\\
  &\Sigma_f(\tau) = -J^2 G_f(\tau)^2G_f(-\tau)+kt^2G_f(\tau)G_b(\tau)G_b(-\tau) \label{f4} \\
  &- G_b (\tau = 0^-) = p \label{f5} \\
 & G_f (\tau = 0^{-}) = \kappa - k p \label{f6}
\end{align}
Thus we have 6 equations to solve for the 6 variables $\overline{r}$, $\mu_f$, $G_f$, $G_b$, $\Sigma_f$, $\Sigma_b$.
Note that Eq. \ref{f1} holds for all $\omega_n$. 

%A Fermi liquid solution will have $\overline{r} \neq 0$ as $\beta \rightarrow \infty$. A non-Fermi liquid solution is expected to have $r \sim \beta^{-a}$ with $0<a<1$ as $\beta \rightarrow \infty$. 

%%%%%%%%%%%%%%%%%%%%%%%%%%%%%%%%%%%%%%%%%%%%%%%%%%%%%%%%%%%%%
\subsection{Real-frequency solution at zero temperature}

\label{sec:realw}

In this section we discuss the solutions of the saddle-point equations on the real-frequency axis at $T=0$. The details of real-frequency equations corresponding to Eqs. \ref{f1}-\ref{f6} can be found in SI Appendix~\ref{sec:REALFL}. 
We look for a low-frequency conformal solution for the fermion and boson Green's functions with the following form:
\begin{equation}
\label{eq:Gfb_c}
    G_{a}(i\omega_n)=-iC_{a}\begin{pmatrix}e^{-i\theta_{a}}\\-e^{i\theta_{a}}\end{pmatrix}\frac{1}{|\omega|^{1-2\Delta_{a}}} \,,
\end{equation}
where the subscript $a=f,b$ corresponds to the fermion and boson Green's function respectively. In the above ansatz, $C_{a}$ is a constant, $\theta_{a}$ is an asymmetry parameter, and $\Delta_{a}$ is the exponent determining the critical solution. Below we shall discuss the relation between these parameters and different possible solutions. The vector notation is introduced to denote the positive and negative frequency parts of the solution. Using this form of the Green's function it is then straightforward to write the corresponding spectral densities,
\begin{align}
    \rho_{a}(\omega)&=-\frac{1}{\pi}\text{Im}[G_{a}(i\omega_n=\omega+i0^+)]\nonumber\\
    &=\frac{C_{a}}{\pi}\begin{pmatrix}\sin(\pi\Delta_a+\theta_a)\\\sin(\pi\Delta_a-\theta_a)\end{pmatrix}\frac{1}{|\omega|^{1-2\Delta_{a}}} \,. 
\end{align}
The exponents of the fermion and boson Green's functions satisfy the constraint $\Delta_f+\Delta_b=1/2$. 
The ansataz considered in Eq. \ref{eq:Gfb_c} admits three types of solutions  \cite{Joshi:2019csz}: \\

(i) \underline{$\Delta_{f} = \Delta_{b} = 1/4$}: 
This is the solution that leads to a marginal spin correlation, i.e., a $1/\tau$ decay as in the SY spin liquid \cite{SY93} found in the insulating case. In  Ref.~\cite{Tikhanovskaya:2020zcw} it was shown that such a solution exists only in a very small doping range near $p=0$. This is also the solution that corresponds to the deconfined critical point discussed in Ref.~\cite{Joshi:2019csz}. We will not discuss this solution any further here, because we believe the actual ground state at very low doping is a spin glass. \\

(ii) \underline{$\Delta_{b}=0 \,, \Delta_{f}=1/2$}: Such a solution would be the analog of the Fermi liquid solution found in the large $M$ limit of Ref. \cite{PG99}. However, it turns out that such a solution is not a valid solution of the saddle-point equations of the large $M$ limit considered here. We provide more details in SI Appendix~1.\ref{sec:fl} in this regard. \\

(iii) \underline{$1/4 < \Delta_{f} < 1/2$}: 
In this solution the $J$ term in the fermion self-energy in Eq. \ref{f4} is sub-dominant at low energies, although it does have contributions at higher energies. This solution results in doping-dependent exponents. As we shall see below, we find this solution to be present for all values of doping, and it will be a valid solution in the overdoped region. Our subsequent analyses in this section will focus on this solution. \\
%and it is consistent with the  imaginary-frequency solution discussed in Section~\ref{sec:imagw}. \\

\begin{figure}[t]
%\centering
    \includegraphics[width=1.01\linewidth]{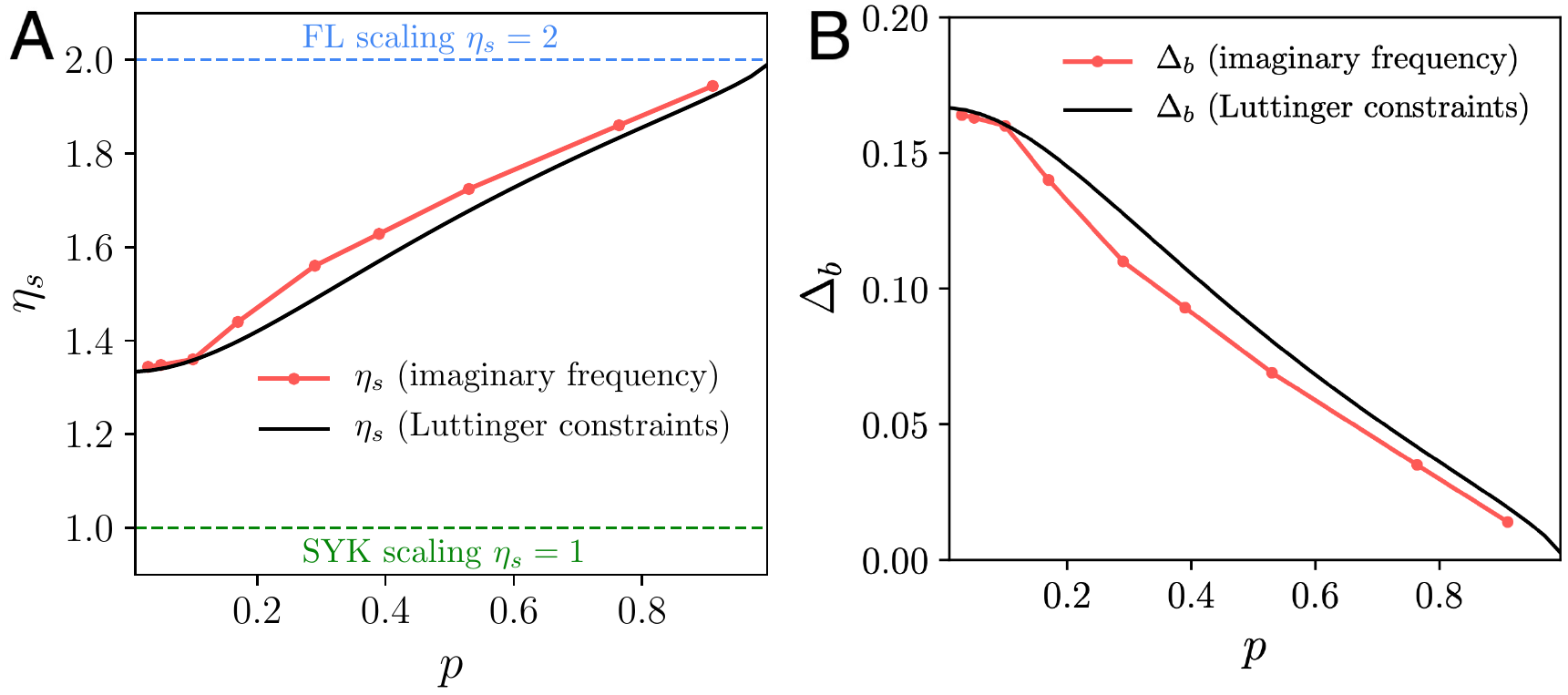}
    \caption{(A) We plot the doping dependence of the exponent $\eta_s$ of the spin correlation $\chi(\tau)\sim 1/|\tau|^{\eta_s}$. We compute $\eta_s=2-4\Delta_{b}$ in two independent ways, one by solving the Luttinger relations and the zero-frequency saddle point equations at $T=0$ as explained in the text (black curve) and by computing $\eta_s$ from the value of $\Delta_b$ obtained from the imaginary-frequency numerics (red curve), as discussed in Section~2.\ref{sec:imagw}. We find the two results in good agreement. At $p=0$ we obtain  $\eta_s=4/3$, with $\eta_s$ increasing monotonically with increasing doping such that $\eta_s \to 2$ as $p \to 1$. (B) We plot the doping dependence of $\Delta_b$ as obtained independently from the luttinger relations and $T=0$ zero-frequency saddle point equations and from solving the imaginary frequency saddle point equations. We find at $p=0$ $\Delta_b=1/6$ and $\Delta_{b} \to 0$ as $p \to 1$. Consequently, this means $\Delta_f=1/3$ at $p=0$ and $\Delta_{f} \to 1/2$ as $p\to 1$.}
    \label{fig:deltab}
\end{figure}

We now briefly discuss our strategy to find the solution (iii) and more details can be found in SI Appendix~\ref{sec:REALFL}. The conformal solution ansataz introduced in Eq. \ref{eq:Gfb_c} satisfies two Luttinger constraints \cite{Joshi:2019csz, Tikhanovskaya:2020zcw}, 
\begin{align}
\label{eq:lutt_con}
\frac{\theta_f}{\pi}+\left(\frac{1}{2}-\Delta_f\right)\frac{\sin(2\theta_f)}{\sin(2\pi\Delta_f)} &= \frac{1}{2}-\kappa+kp \,, \nonumber \\
\frac{\theta_b}{\pi}+\left(\frac{1}{2}-\Delta_b\right)\frac{\sin(2\theta_b)}{\sin(2\pi\Delta_b)} &= \frac{1}{2}+p \,.
\end{align}
For solution (iii) the constants $C_{f}$ and $C_{b}$ can not be determined individually but their product is a constant. This leads to another constraint involving $\theta's$ and $\Delta's$ as shown in Ref.~\cite{Joshi:2019csz}, 
\begin{align}
\label{eq:con3}
k=-\frac{\Gamma(2-2\Delta_f)\Gamma(2\Delta_f)\sin(\pi\Delta_f+\theta_f)\sin(\pi\Delta_f-\theta_f)}{\Gamma(2-2\Delta_b)\Gamma(2\Delta_b)\sin(\pi\Delta_b+\theta_b)\sin(\pi\Delta_b-\theta_b)} \,,
\end{align}
where $k=1/2$ in our case. Together with the constraint $\Delta_f + \Delta_b = 1/2$ and Eqs. \ref{eq:lutt_con}-\ref{eq:con3}, there are four equations to solve for four variables, namely $\Delta_f$, $\Delta_b$, $\theta_f$, and $\theta_b$, at a fixed doping $p$. Solving these equations gives us these parameters as a function of doping, as shown in Fig.~\ref{fig:deltab} and Fig.~\ref{fig:vals}. Determination of the constants $C_f$ and $C_b$ requires full solution of the saddle point equation at all frequencies, and the resulting values are shown in Fig.~\ref{fig:vals}. 
Of particular interest is the doping dependence of $\Delta_b$, shown in Fig.~\ref{fig:deltab}. 

In the large-$M$ limit, from Eq. \ref{Eq:EoM5} it is clear that the anomalous dimension of the electron operator is $\eta_c = 2 (\Delta_f + \Delta_b) = 1$ and that of the spin operator is $\eta_S = 4\Delta_{f}=2-4\Delta_{b}$. Therefore, from Eq. \ref{Eq:EoM5}, as a function of the imaginary time the electron Green's function and the local spin correlation have the form, 
\begin{equation}
\label{eq:Gc_Gs}
G_{c}(\tau) \sim \frac{1}{\tau} \,, ~~~~
\chi(\tau) \sim \frac{1}{\tau^{2-4\Delta_{b}}} \,.
\end{equation}
Thus we find that although the electron Green's function is Fermi-liquid like the spin correlation is not. Therefore the solution we have found is a critical metallic phase with a doping-dependent exponent of the spin correlation. Only in the limit $p\to1$ we have $\Delta_b \to 0$, leading to a Fermi-liquid like spin correction with a $1/\tau^{2}$ decay.

\begin{figure}[t]
    \centering
    \includegraphics[width=.7\linewidth]{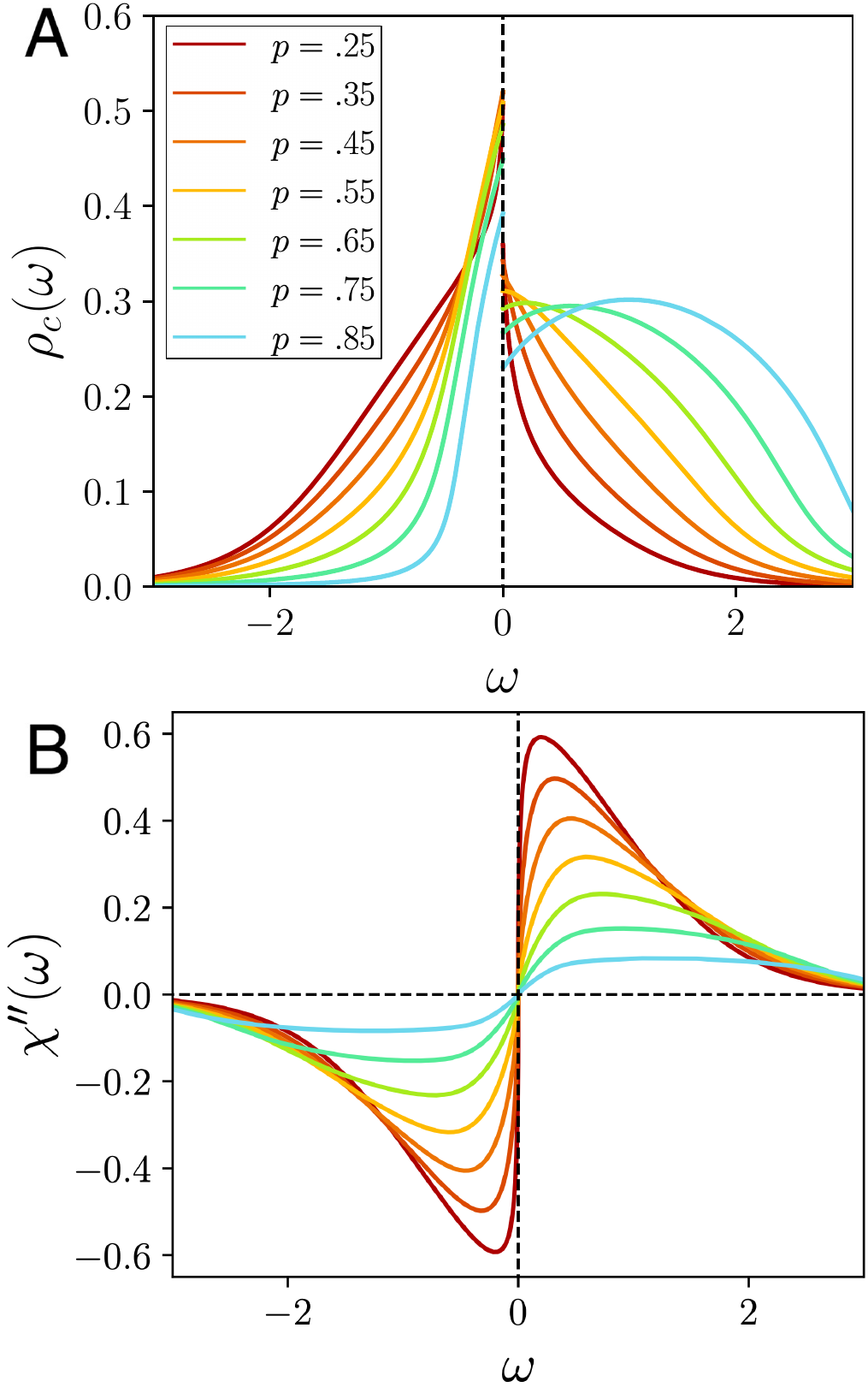}
    \caption{Plot of the electron spectral density (left) and spin spectral density (right) obtained at $t=J=1$. The electron spectral density which discontinuously approaches two different constants $\omega=0$, leading to $1/\tau$ decay of the electron density in imaginary time. The spin spectral density goes as $\text{sgn}(\omega)|\omega|^{4\Delta_f-1}$ as $\omega\rightarrow 0$. Only for large $p$ do we have $\Delta_f$ approach 1/4, which implies a linear frequency dependence and $1/\tau^{2}$ decay behavior in imaginary time characteristic of a Fermi liquid.}
    \label{fig:re_rs}
\end{figure}

Having established the presence of solution (iii), we now solve the saddle-point equations on the real-frequency axis numerically to obtain the full $\om$ dependence of the boson and fermion spectral densities. Using these we can also obtain the electron and spin spectral densities. 
Fig.~\ref{fig:rf_rb} shows the full $\omega$-dependent solution for the fermion and boson spectral densities, while in Fig.~\ref{fig:re_rs} we plot the electron and spin spectral densities. The boson and fermion spectral densities have the form $\rho_{a}(\om) \sim \om^{2\Delta_{a} - 1}$ with $a=f,b$. Since $\Delta_{a}<1/2$, we plot the rescaled spectral densities in Fig.~\ref{fig:rf_rb}. 
Note that the fermion and boson spectral densities are not observable quantities. However, the electron and spin spectral densities are observable in photoemission and neutron scattering experiments. These are defined as follows:
\begin{align}
\label{eq:spec_den}
&\rho_{c} = -\frac{1}{\pi} \mbox{Im} [G_{c}(\iw \to \om + i0^{+})] \\
&\rho_{s} = \frac{1}{\pi} \mbox{Im} [\chi(\iw \to \om + i0^{+})] = \frac{1}{\pi} \chi''(\om) \,.
\end{align}
In SI Appendix \ref{sec:REALFL} we present more details on evaluating these spectral densities.

As noted above, the most striking feature of our solution is a continuously varying spin-correlation exponent. This is clearly seen in Fig.~\ref{fig:deltab}\textit{A} which shows the exponent of the spin correlator $\chi(\tau)\sim 1/|\tau|^{\eta_s}$ where $\eta_s=4\Delta_f$, and in Fig.~\ref{fig:re_rs}\textit{B}, where $\chi''(\om) \sim \text{sgn}(\om)|\om|^{4\Delta_{f} -1}$ at low frequencies. Only in the limit $p\to 1$, we see a Fermi-liquid like behavior, $\chi''(\om) \sim \om$, as $\Delta_{f} \to 1/2$. The electron spectral density is a constant in the low-frequency limit with different values for $\om \to 0^{+}$ and $\om \to 0^{-}$ (see Fig.~\ref{fig:re_rs}\textit{A}), thus resulting in a discontinuity at $\om = 0$. It clearly displays a particle-hole asymmetry throughout this phase, which is relevant in the context of understanding the measurement of Seebeck coefficient in recent experiments \cite{gourgout2021seebeck, Georges2021seebeck}. 

As for the SYK model \cite{syk_rmp}, the conformal solutions to Eqs. \ref{Eq:EoM1}-\ref{Eq:EoM4} have time reparameterization symmetry when the self-energies are singular so that $G_{f,b} (i \omega_n) \Sigma_{f,b} (i \omega_n) \approx -1$. Ref.~\cite{Guo:2020aog} discussed a mechanism relating the time reparameterization symmetry to a linear-$T$ contribution to the resistivity, and the same mechanism applies to the cases (i) and (iii) above. Briefly, to obtain a model with spatial structure, we consider the $t$-$J$ model on a large dimension lattice with non-random $t_{ij}$ but random nearest-neighbor $J_{ij}$. In the limit of large dimension, the self-energies become local, and the Green's functions and self-energies obey equations closely related to Eqs.~\ref{Eq:EoM1}-\ref{Eq:EoM4}. The conductivity in this large dimension model is given by the Kubo formula applied at one loop to the electron Green's function. The identity $\Delta_f + \Delta_b = 1/2$ implies that $G_c \sim 1/\tau$, and inserting this leading scaling behavior into the Kubo formula leads to a $T$-independent residual resistivity. To obtain temperature dependence, we consider the corrections to scaling from the time reparameterization operator, whose scaling dimension $h=2$ leads to corrections which depend linearly on $T$ or $\omega$. Applying such a correction to the residual resistivity, we obtain a linear-in-$T$ resistivity.
%Experiments on overdoped cuprates have reported a linear-$T$ resistivity in an extended region beyond the putative quantum critical point \cite{Cooper09, Hussey21}. 
The critical metal phase found here is therefore an attractive candidate for the overdoped cuprates. 
We also note that our solution is consistent with the findings of recent numerical work on a similar model \cite{Dumitrescu2021}, as we will discuss further in Section~\ref{sec:disc}.

%our model and it may be a property of the large-$M$ saddle-point equations in this random and all-to-all interactions model. However, at finite $M$ there may exist a Fermi-liquid solution. Such an investigation goes beyond the scope of present work. We also note that a lattice model with random hopping and exchange interactions between nearest neighbors does lead to a Fermi-liquid phase \cite{PG99}. This results from fully condensing the bosonic holons. However, such an ansatz leads to a Fermi-liquid solution down to infinitesimally small doping values. A possible remedy to it would be to consider an ansatz with condensate contribution as well as a conformal part. However, we find that such a solution is not consistent and discuss some details in Appendix \ref{sec:fl}. 

%%%%%%%%%%%%%%%%%%%%%%%%%%%%%%

%%%%%%%%%%%%%%%%%%%%%%%%%%%%%%

%%%%%%%%%%%%%%%%%%%%%%%%%%%%%%
%%%%%%%%%%%%%%%%%%%%%%%%%%%%%%

%%%%%%%%%%%%%%%%%%%%%%%%%%%%%%
%%%%%%%%%%%%%%%%%%%%%%%%%%%%%%

%%%%%%%%%%%%%%%%%%%%%%%%%%%%%%%%%%%%%%%%%%%%%%%%%%%%%%%%%%%%%%%%
\subsection{Imaginary-frequency solution}
\label{sec:imagw}

\begin{figure}
    \centering
    {\includegraphics[width=.8\linewidth]{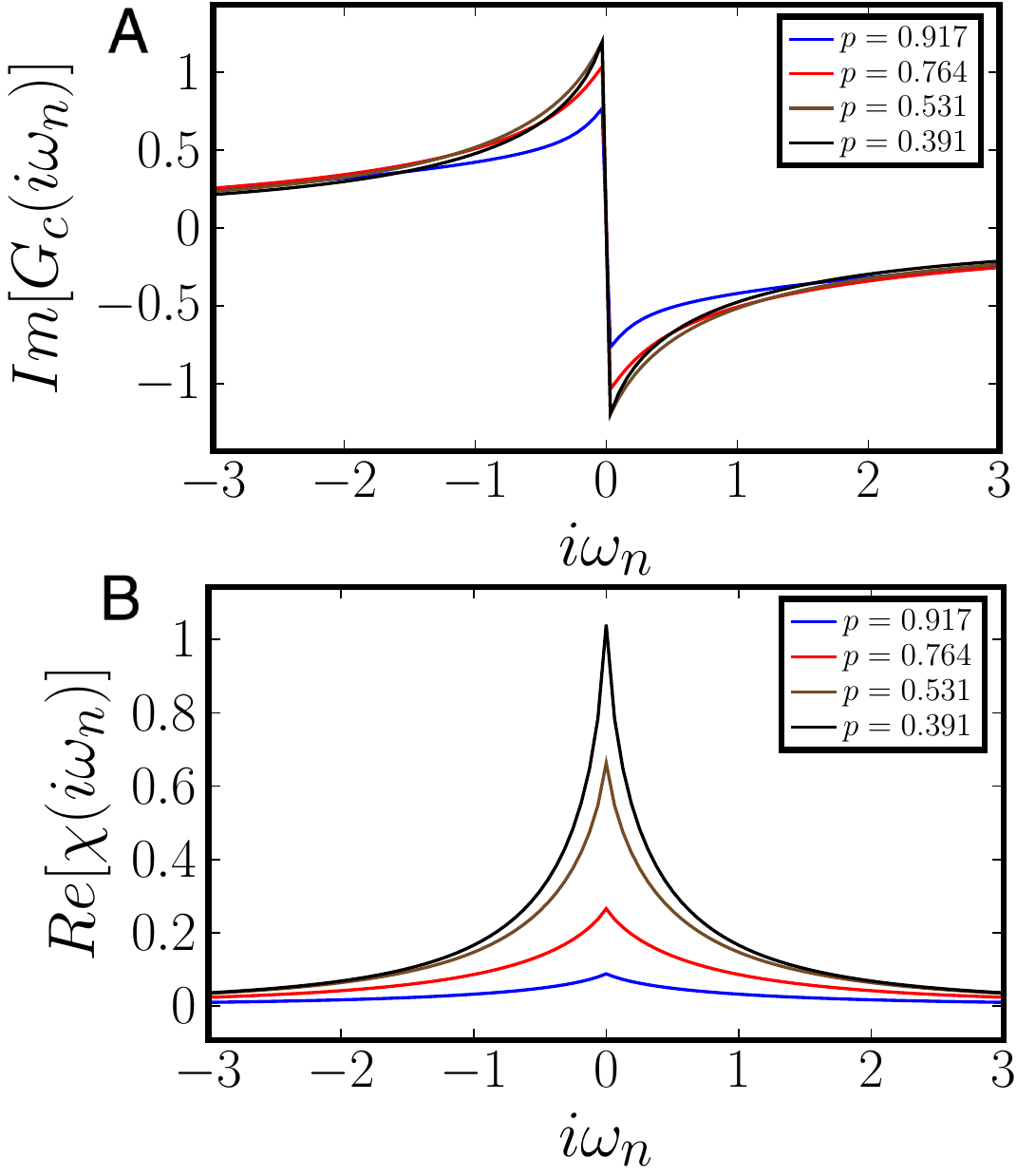}} \\
% ~~~  {\includegraphics[width=.75\linewidth]{./RGsw}}
    \caption{(A) Imaginary part of the electron Green's function on the Matsubara-frequency axis at $T=0.01$ for different dopings. (B) Real part of the spin correlation on the Matsubara-frequency axis at $T=0.01$ for different dopings. We have used $t=J=1$.}
    \label{fig:GcGsw}
\end{figure}

We also numerically solve Eqs. \ref{f1}-\ref{f6} on the imaginary-frequency axis at finite temperatures and for different doping values. A critical metallic solution is found for all values of doping. We calculate the parton Green's functions as well as gauge-invariant observables, namely, electron Green's function and spin correlation. These are shown in Fig.~\ref{fig:GcGsw}.
The exponent of the bosonic Green's function, $\Delta_b$, introduced in Eq. \ref{eq:Gfb_c} can be determined from the temperature dependence of $G_{b}(\iw=0)$. In the low-temperature limit, the bosonic Green's function of the critical metallic solution has a conformal form at low frequencies, which follows the relation
\begin{equation}
\label{eq:Gb0_T}
G_{b}(\iw=0) = C_{0} T^{-1 + 2\Delta_{b}} \,,
\end{equation}
where $C_{0}$ is some constant and $T$ is the temperature. In order to extract $\Delta_b$ from the data, we plot
$\log(G_{b}(\iw=0))$ as a function of $\log(T)$ and perform a linear fit. The slope of this linear fit then determines $\Delta_b$. In Fig.~\ref{fig:deltab} we plot $\Delta_b$ determined from this procedure as a function of doping. It is in good agreement with the result obtained from analytic solution (shown as black curve in Fig.~\ref{fig:deltab}) discussed in Section~2.\ref{sec:realw}. The lowest temperature that we used for the procedure is $T=0.01$. 

In the SI Appendix~\ref{sec:imag_app} we present additional results obtained from solving the saddle-point equations on the imaginary-frequency axis. In particular, we calculate the electron Green's function and the spin correlation, which are physical observables. We find that the spin correlation $\chi(\tau)$ fits the conformal form to a very high accuracy and this allows us to extract the corresponding exponent. Moreover, using Pade approximation, we also perform a numerical analytic continuation to the real-frequency axis to obtain the electron and spin spectral densities. These are discussed in SI Appendix~2.\ref{sec:nac}, and are in remarkable agreement with those obtained from real-frequency analysis at zero temperature. 

%%%%%%%%%%%%%%%%%%%%%%%%%%%%%%%%%%%%%
%%%%%%%%%%%%%%%%%%%%%%%%%%%%%%%%%%%%%

%%%%%%%%%%%%%%%%%%%%%%%%%%%%%%%%%%%%%%%%%%%%%%%%%%%%%%%%%%%%%%%
\subsection{Instability to spin-glass phase/quantum critical point}
\label{sec:qcp}

\begin{figure}[t]
  %  \centering
    \includegraphics[width=1.02\linewidth]{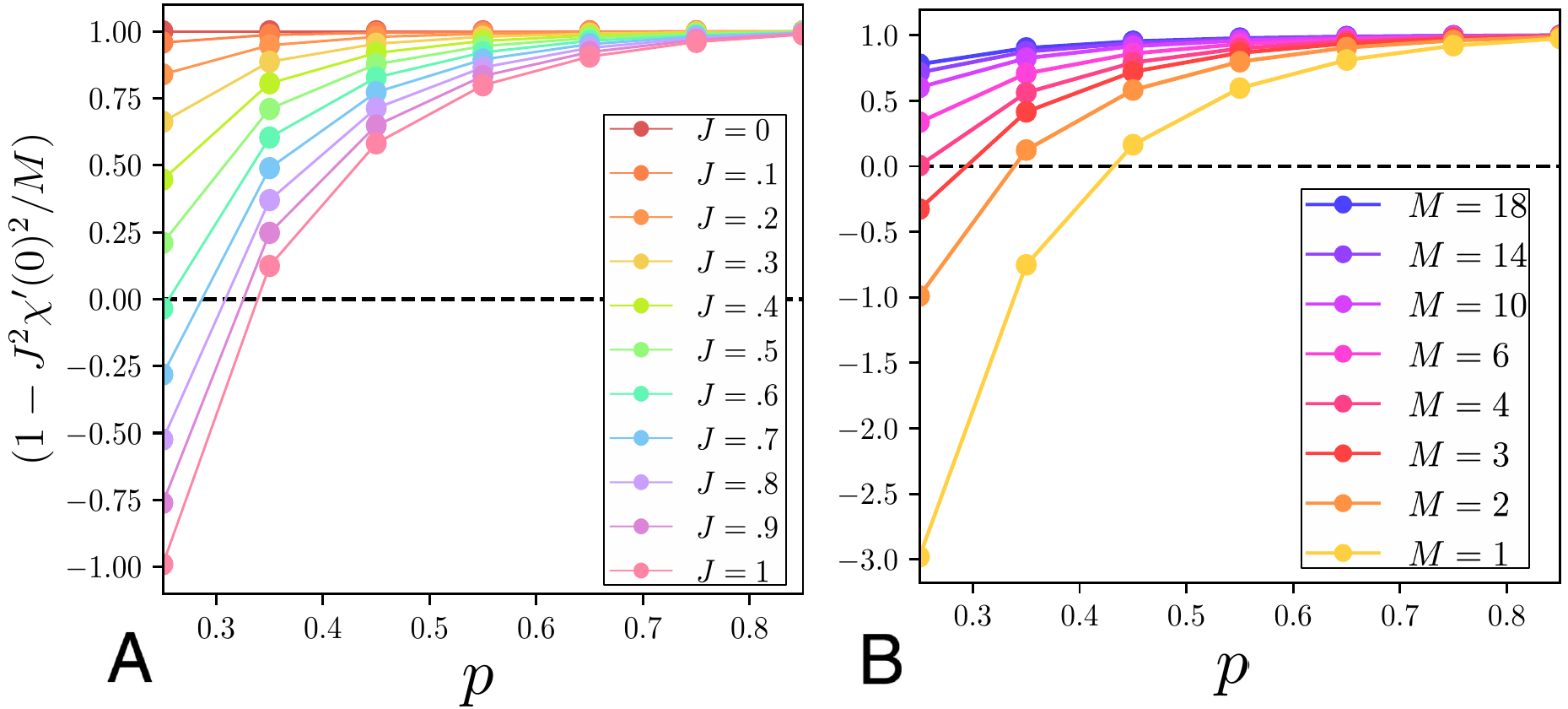}
    \caption{ Plots of the coefficient of the quadratic term in the free energy in Eq. \ref{eq:fe} as a function of doping for (A) different values of $J$ with $t=1$ and $M=2$; 
    (B) different values of $M$ with $t=J=1$. In both plots, $T=0$.}
    \label{fig:susceptibleJ}
\end{figure}

\begin{figure}
    \centering
    \includegraphics[width=.75\linewidth]{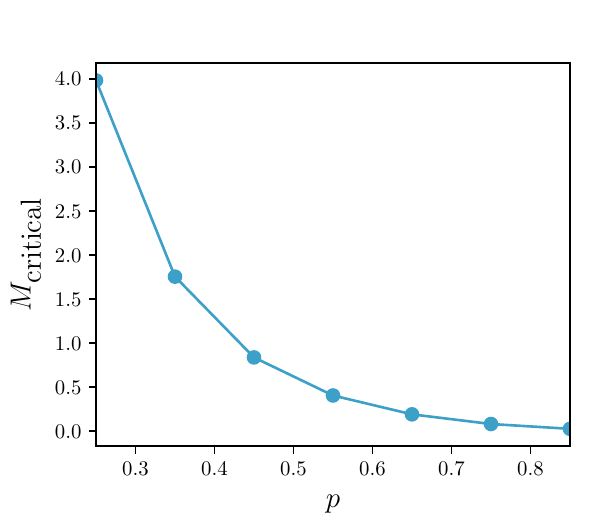}
    \caption{Plot showing critical value of $M$ for which coefficient of quadratic term in free energy Eq. \ref{eq:fe} changes sign at zero temperature. }
    \label{fig:criticalm}
\end{figure}

The critical metallic solution discussed above is expected to be stable at large doping values. In the low doping region we expect a spin-glass phase, which is connected to the spin-glass phase found in the insulating case \cite{Christos2021}. The critical metallic phase and the spin-glass phase are separated by a quantum critical point at a finite doping. The critical value of doping can be estimated using a Ginzburg-Landau type free energy considered in Ref.~\cite{Christos2021}. It was derived by considering fluctuations over the saddle point leading to a $1/M$ correction and it has a form,
\begin{equation}
\label{eq:fe}
\mathcal{F} = \left[ 1 - \frac{J^{2}}{M} \chi^{2}(i\om=0) \right] \Psi^{2} + \ldots \,,
\end{equation}
where $\chi(i\om=0)=\int_{0}^{\beta} \chi(\tau) d\tau$ is the local spin susceptiblity and $\Psi$ is the spin-glass order parameter. 
In the insulating $p=0$ phase studied in Ref.~\cite{Christos2021}, it was found that $\chi(i \omega =0 )$ diverged logarithmically at $T=0$ because the spin exponent has the `marginal' value $\eta_s = 1$. Consequently the co-efficient of $\Psi^2$ is always negative as  $T \rightarrow 0$, and spin glass order is present at $T=0$. In our non-Fermi liquid solution, $\eta_s < 1$, and so $\chi(i \omega =0)$ is finite at $T=0$: this allows the possibility of $\Psi=0$ and no spin glass order at $T=0$. 
We find in the overdoped region the coefficient of the quadratic term is positive and the free energy is thus minimized by  $\Psi=0$. On the other hand, in the underdoped region this coefficient is negative leading to a non-zero spin-glass order parameter. A change in the sign of the coefficient of $\Psi^{2}$ at $T=0$ thus indicates a quantum phase transition to a spin-glass phase and can be used to estimate the critical value of doping. We plot this coefficient at zero temperature in Fig.~\ref{fig:susceptibleJ}\textit{A} as a function of doping for different values of $J$ at $M=2$. We clearly see that for larger values of $J$ there is a quantum phase transition at a finite doping. In particular, for $J=1$ we find $p_{c} \approx 0.33$ and for $J=0.5$ we get $p_c \approx 0.25$. Similarly, in Fig.~\ref{fig:susceptibleJ}\textit{B} we plot the coefficient in Eq. \ref{eq:fe} as a function of doping for different values of $M$ at a fixed $J=1$ at zero temperature. It is clear that for large values of $M$ there is no phase transition. We also plot the critical value of $M$ as a function of doping in Fig.~\ref{fig:criticalm} at zero temperature. We perform a similar analysis at finite temperature to obtain the critical doping as a function of temperature (see Fig.~\ref{fig:c_p}). The resulting phase diagram is shown in Fig.~\ref{fig:pd}\textit{A}. The spin-glass susceptibility increases with decreasing temperature and thus the spin-glass phase is found upon {\em cooling} the non-Fermi liquid at low doping. However, note that unlike in the case of the random Heisenberg model \cite{SY93, GPS01} (where the solution $(i)$ with $\Delta_f = \Delta_b =1/4$ is present) the spin-glass susceptibility does not diverge at low temperature in the present case. This is one of the important reasons for a stable critical solution at zero temperature in a broad doping region. We also note that while $1/M$ corrections are used in deriving the form of Eq. \ref{eq:fe}, we have used the $M\rightarrow\infty$ solution for the critical metal to compute $\chi'(0)$ as it appears in Eq. \ref{eq:fe}. While a more accurate estimate of $p_c$ may be obtained by adding $1/M$ corrections to the critical metal solution, we emphasize that the $M\rightarrow\infty$ solution is sufficient to capture existence of a finite doping phase transition between the spin glass and critical metal.

%In Ref. \cite{Joshi:2019csz}, using renormalization group (RG) technique a deconfined critical point was proposed, separating the spin-glass and Fermi-liquid phases. It had the feature that the spin correlation was exactly marginal, just like in the SYK models \cite{SY93, kitaev2015talk}. The critical point that we find here does not have a marginal spin correlation and the over-doped phase is not a Fermi liquid but a critical metal. One possibility is that the large-$M$ solution found here may lead to a marginal behavior at the critical point at finite $M$ once corrections beyond saddle point are taken into account. Another possibility is that the stable fixed point found in Ref. \cite{Joshi:2019csz} may be unstable at strong coupling to another fixed point. In fact, in Ref. \cite{Joshi:2019csz} there exists a fixed point with the property of $\eta_c =1$ and varying $\eta_S$, which may become stable at strong coupling. 

%%%%%%%%%%%%%%%%%%%%%%%%%%%%%%
%%%%%%%%%%%%%%%%%%%%%%%%%%%%%%

%%%%%%%%%%%%%%%%%%%%%%%%%%%%%%
%%%%%%%%%%%%%%%%%%%%%%%%%%%%%%

%%%%%%%%%%%%%%%%%%%%%%%%%%%%%%%%%%%%%%%%%%%%%%%%%%%%%%%%%%%%%%%%
\section{Spin-glass phase}
\label{sec:underd}

In this section we discuss the spin-glass phase present at lower dopings. In this case, we shall use the representation where spinons are bosonic ($\bs$) and holon is a fermion operator ($\fh$). In terms of these operators, 
\begin{equation}
\label{eq:cs_sg}
c_{\alpha} = \fh^{\dagger} \bs_{\alpha} \,, ~~~~
\vec{S} = \bs^{\dagger}_{\alpha} \frac{\vec{\sigma}_{\alpha\beta}}{2} \bs_{\beta} \,,
\end{equation}
and we realize a SU$(2|1)$ superalgebra.
Just as before, we shall now enlarge the symmetry here to SU$(M|M')$, which means $\alpha=1,\ldots,M$ and the holon operator acquires an index $\ell=1,\ldots,M'$. Note that this bosonic spinon large $M$ limit is {\it distinct\/} from the fermionic spinon large $M$ limit followed in Section~\ref{sec:overd}, and the two models are the same {\it only\/} for $M=2$; so there is no reason to expect quantitive agreement  between the results of the present section and those of Section~\ref{sec:underd}.

\begin{figure}
    \centering
    \includegraphics[width=0.8\linewidth]{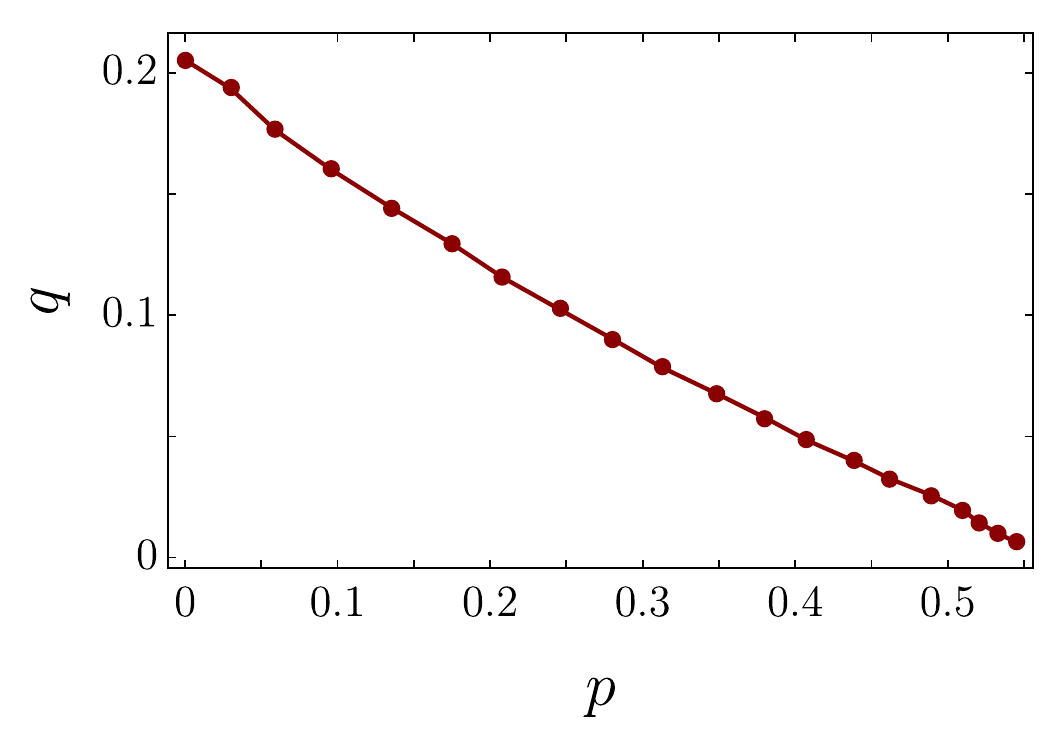}
    \caption{Plot of spin-glass order parameter, $q=g^{2}$, as a function of doping obtained by solving the bosonic spinon saddle-point equations (\ref{eq:sg_Gr} - \ref{eq:sg_p}) on the real-frequency axis at $T=0$. The point at $p=0$ is obtained by solving equations (\ref{ysr14a}-\ref{ysr14c}) at zero temperature on the real frequency axis. The parameters of the model are $J=t=1$. The critical doping $p_c$ at which the spin glass order vanishes in the present bosonic spinon computation need not be the same as that in the fermionic spinon computation in Section~2.\ref{sec:qcp} because the models are identical only for $M=2$.}
    \label{fig:sg_op}
\end{figure}

The strategy to obtain the saddle-point equations is similar to that discussed earlier in Section~\ref{sec:overd}. From Refs.~\cite{Joshi:2019csz,Tikhanovskaya:2020zcw}, we have
\begin{align}\nonumber
  & G_{\bs} (i\omega_n) = \frac{1}{i\omega_n+\mu_{\bs} -\Sigma_{\bs}(i\omega_n)}\,, \\\nonumber
  &G_{\fh} (i\omega_n) = \frac{1}{i\omega_n+\mu_{\fh}-\Sigma_{\fh}(i\omega_n)}\,, \\\nonumber
  &\Sigma_\bs (\tau) = -k t^2 G_\fh(\tau)G_\fh(-\tau)G_\bs(\tau)+J^2 G_\bs(\tau)^2 G_\bs(-\tau)\,, \\\nonumber
  &\Sigma_\fh (\tau) = t^2G_\fh(\tau)G_\bs(-\tau)G_\bs(\tau)\,, \\\nonumber
  &G_\bs (\tau=0^-) =- \kappa + k p  \,, \\\nonumber
  &G_\fh (\tau = 0^-) = p  \,.
\end{align}
Physically, the spin-glass phase can be understood as a condensation of bosonic spinons.
Contrary to the earlier case, since we are interested in the spin-glass phase we need to retain the replica off-diagonal terms in the action upon disorder averaging. There is however some simplification and only replica off-diagonal components of the bosonic Green's function at $i\om_{n}=0$ are relevant. These are captured via the parameter $g$ which is related to the spin-glass order parameter, as discussed in detail in SI Appendix~\ref{sec:SY}. After incorporating the spin-glass order in $G_\bs(\tau) = G_{r}(\tau)-g$, we obtain the following saddle-point equations: 
\begin{align}
\label{eq:sg_Gr}
&\left[ G_r (i \omega_n) \right]^{-1} = i \omega_n - \frac{Jg}{\Theta} - \left[ \Sigma_r (i \omega_n) - \Sigma_r (i \omega_n = 0) \right]  \\
 &G_\fh (i\omega_n) = \frac{1}{i\omega_n+\mu_\fh-\Sigma_\fh(i\omega_n)}\,, \\
 &\Sigma_r (\tau) = J^2 \Bigl( \left[G_r (\tau) \right]^2 G_r (-\tau)  - 2 g G_r (\tau) G_r (-\tau) - g \left[G_r (\tau) \right]^2  \nonumber \\
&+ 2 g^2 G_r (\tau) + g^2 G_r(-\tau) \Bigr)  -k t^2 G_\fh(\tau)G_\fh(-\tau) \Bigl( - g + G_r(\tau) \Bigr),\\
 &\Sigma_\fh (\tau) = t^2G_\fh(\tau)\Bigl( g^2 - g G_r (\tau) - g G_r(-\tau) + G_r(-\tau)G_r(\tau) \Bigr)\,, \\
 &g - G_r (\tau = 0^-) = \kappa - k p \\
 &G_\fh (\tau = 0^-) = p \,. \label{eq:sg_p}
\end{align}
The dimensionless parameter $\Theta=1/\sqrt{3}$, as detailed in SI Appendix~\ref{sec:SY}.
We solve the above equations on real and imaginary-frequency axes. The main observable in this phase is the spin-glass order parameter, which we plot as a function of doping in Fig.~\ref{fig:sg_op} at zero temperature. The spin-glass order parameter is finite at lower dopings and decreases upon increasing doping. We show results for the order parameter computed for small but finite temperature obtained using the imaginary frequency analysis in SI Appendix~\ref{sec:imag_app_sg}. We also calculate the holon and spinon Green's functions as well as the electron Green's function and spin correlation at finite temperature. These quantities are detailed in the same SI Appendix~\ref{sec:imag_app_sg}. 

\begin{figure}[t]
    \centering
    {\includegraphics[width=.8\linewidth]{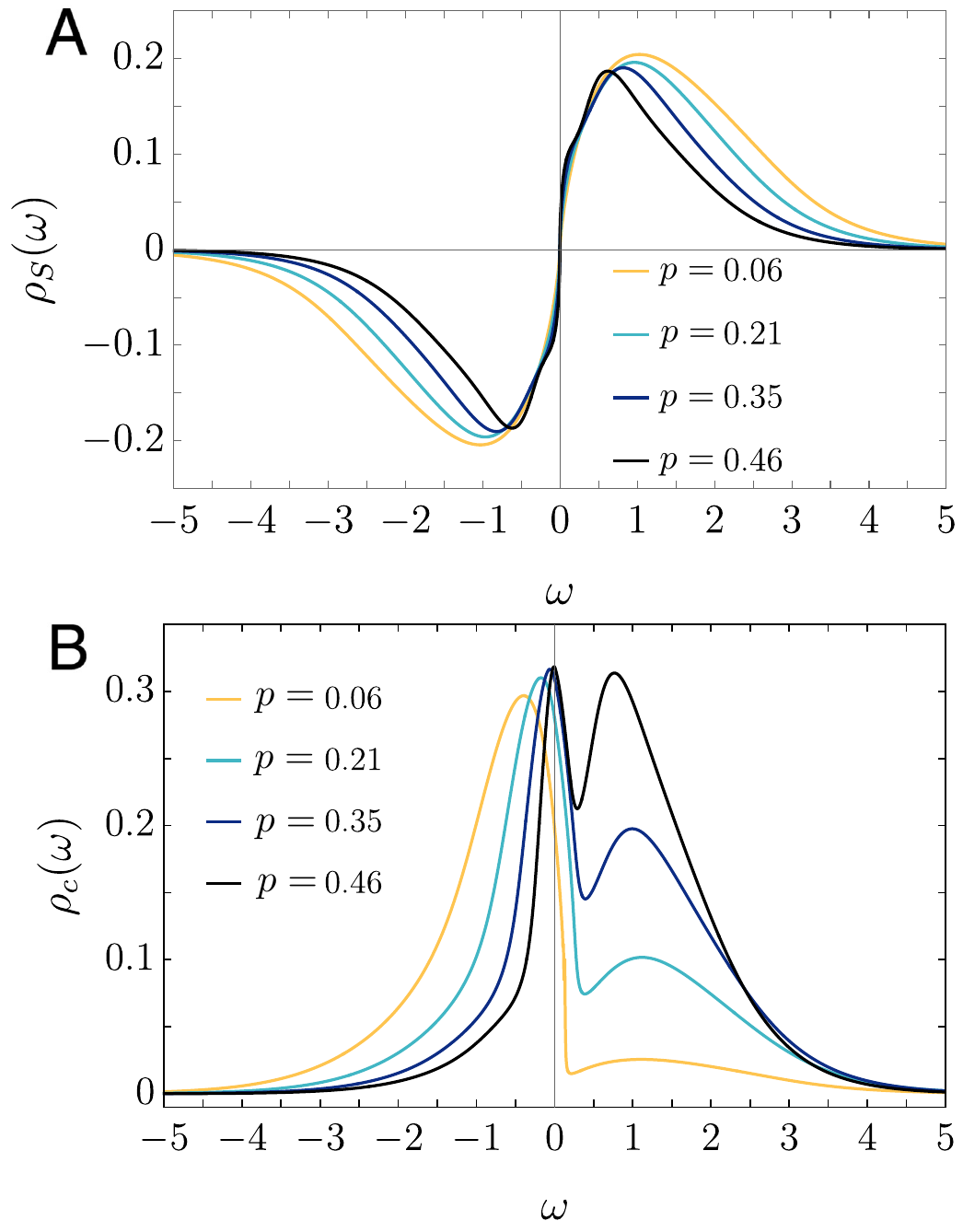}} 
    %{\includegraphics[width=.8\linewidth]{./rhoel_sg_T0}}
    \caption{ (A) Spin spectral function and (B) electron spectral function obtained at zero temperature via real frequency analysis for several values of doping. Note the linear in frequency behavior at small frequencies for $\rho_S(\omega)$. Parameters are $J=t=1$.}
    \label{fig:rhoSG}
\end{figure}

We solve the saddle-point equations Eqs. \ref{eq:sg_Gr}-\ref{eq:sg_p} at zero temperature to obtain the fermion and boson spectral functions, as well as the electron and spin spectral functions for a range of doping. In terms of the spinon and holon spectral functions the electron spectral function has the following form: 
\bea
\rho_c(\omega) =\int_{0}^{\omega}d\omega_1\rho_r(\omega_1)\rho_\fh(\omega_1-\omega)+ g\rho_\fh(-\omega) \,.
\eea
Similarly, we obtain the expression for the spin spectral function, 
\begin{align}
\label{eq:rs_sg}
\rs(\om)  
&= g^{2} \beta \om \delta(\om) + g \left[ \rho_{r}(\om) - \rho_{r}(-\om) \right]\nonumber \\
&- \int_{0}^{\om} d\om_{1} ~\rho_{r}(\om_{1}) \rho_{r}(\om-\om_{1}) \,,
\end{align}
as in Ref.~\cite{GPS01}. The spectral functions at zero temperature for different values of doping are shown in Fig.~\ref{fig:rhoSG}. We note that the boson spectral function behaves linearly with frequency at small frequencies. 
Consequently, the spin spectral function also depends linearly on $\om$ at small frequencies. 
%which is what is expected in the spin glass phase. 
We also note the double peak structure in the electron spectral function $\rho_c(\omega)$ in Fig.~\ref{fig:rhoSG}\textit{B}. 
%We believe the structure is coming from the boson spectral weight $\rho_{\bs}(\omega)$ dominated at negative frequencies and fermions $\rho_\fh(\omega)$ at positive frequencies. 
Recall that the electron spectral function is a convolution of the spinon and holon spectral functions, as well as a term proportional to $g \rho_{\fh}(-\om)$.
At small dopings the value of $g$ is larger and hence there is a dominant peak coming from fermion spectral function. As $g$ decreases with increasing doping the contribution from the boson spectral function increases leading to the second peak at positive frequencies. 
%At small dopings the electron spectral weight is primarily dominated by the spinon spectral function which gives a dominant peak boson contribution and as the doping increases, the fermion spectral weight becomes more important. 
%Notice the linear behavior for small frequencies in the spin spectral function $\rho_S(\omega)$ in Fig.~\ref{fig:rhoSG}(c). 
All the spectral functions satisfy the respective sum rules, which are detailed in SI Appendix~4.\ref{sec:sg_sd}.

Apart from the numerical analysis at zero temperature, we also perform a Pade-approximation based numerical analytic continuation of the imaginary-axis solution obtained at finite temperature. We evaluate the holon, spinon, electron, and spin spectral densities at finite temperature. These are discussed in SI Appendix~2.\ref{sec:nac}, and are consistent with those obtained from real-frequency calculation at zero temperature for some range of doping. 

%%%%%%%%%%%%%%%%%%%%%%%%%%%%%%%%%%%%%%%%%%%%%
%%%%%%%%%%%%%%%%%%%%%%%%%%%%%%%%%%%%%%%%%%%%%

%%%%%%%%%%%%%%%%%%%%%%%%%%%%%%%%%%%%%%%%%%%%%%%%%%%%%%%%%%%%%%%%%%%%%%%%%%%%%
\section{Discussion}
\label{sec:disc}

We have presented a large-$M$ solution for the random $t$-$J$ model for the entire doping range. Our main finding is a critical non-Fermi-liquid metal phase at large dopings. This phase is characterized by a spin correlation exponent which varies continuously with doping, a linear in temperature contribution to the resistivity as $T\to 0$, and an electron spectral function with a Fermi-liquid-like decay at long time, but with a pronounced particle-hole asymmetry. This critical phase captures many aspects of experimental observations in the overdoped cuprates. It has been observed that in the overdoped region of cuprate materials there is a broad range of doping where the resistivity display a linear-$T$ behavior \cite{Cooper09,Hussey21}. Our findings propose a possible mechanism for this observation. Also, recent Seebeck coefficient measurements \cite{gourgout2021seebeck,Georges2021seebeck} hint towards a particle-hole asymmetric electron spectral density, which our solution also displays. 
It turns out that in our solution the electron Green's function appears Fermi-liquid like although the spin correlation does not. This may also explain the fact that experiments on overdoped cuprates probing electron Green's function directly, such as photoemission \cite{arpes_chen2019}, may observe a Fermi-liquid behavior and can not access the critical phase. However, transport measurements obtain properties such as linear-$T$ resistivity which is starkly in contrast to a Fermi liquid. 

We also show that the overdoped critical phase has an instability towards a spin-glass phase at lower dopings. The spin-glass phase is characterized by a spin-glass order parameter, which we calculate using bosonic spinons. We show that this order parameter decreases upon increasing doping. In the context of cuprates, recent experiments have reported the presence of spin-glass phase at low doping \cite{Frachet2020}. Our work therefore presents a comprehensive analysis of model in Eq. \ref{eq:Ham} at variable doping and captures the quantum phase transition between the spin-glass phase and a critical non-Fermi-liquid metal. We also note that our results are consistent with recent numerical work on a related model \cite{Dumitrescu2021}. 

A notable feature of our results is that we never find a Fermi-liquid phase in our large $M$ limit of the $t$-$J$ model, as discussed in SI Appendix~1.\ref{sec:fl}. This is in contrast to the distinct large $M$ limit of Ref.~\cite{PG99}, in which the boson $b$ condenses at all non-zero $p$, leading to a Fermi-liquid phase at all doping. 
The two large $M$ limits co-incide only for the physical value $M=2$, and it is an open question which large $M$ limit yields the correct picture at $M=2$ for the random $t$-$J$ model. However, we will note that the numerical study of the $M=2$ case in Ref.~\cite{Dumitrescu2021} does show indications of non-Fermi liquid behavior in the overdoped regime over the $T$ range studied, as their measured spin exponent $\eta_s$ varies with doping in a manner consistent with Fig.~\ref{fig:deltab}\textit{A}.
Thus, although a Fermi liquid phase may eventually appear at very low $T$ in the overdoped regime
for $M=2$, it does appear that our non-Fermi liquid solution is an attractive description of the physics over a wide range of temperatures and dopings accessible in numerics and experiments.

%Note that in our present analysis we do not find a Fermi-liquid phase and we have discussed its reasons and possibilities earlier. However, in a realistic lattice model a Fermi-liquid is expected to be present. For the overdoped cuprates it may not be wrong to postulate that beyond the putative quantum critical point in a broad region of doping there may exist a critical non-Fermi-liquid metal analogous to that discussed here, which then crosses over into a Fermi-liquid metal at higher dopings. 

In conclusion, our work presents a critical metallic phase as an attractive candidate for the overdoped cuprates which matches observations over a significant temperature range. This phase is obtained for a model with random and all-to-all interactions. Although such a model is far from the microscopic Hamiltonian of the cuprates,
the saddle point equations solved are closely related to dynamic mean field equations of more realistic models in finite dimensions \cite{Haule1,Haule2}, as has been extensively discussed in a recent review \cite{syk_rmp}. At very low temperatures, we ultimately expect a crossover to behavior characteristic of finite-dimensional systems, and describing this crossover remains an important topic for future research.

%%%%%%%%%%%%%%%%%%%%%%%%%%%%%%%%%%%%%%%%%%%%%%%%%%%%%%%%%%%%%%%%%%%%%%%%%%%%%
\subsection*{Acknowledgements}

This research was supported by the  U.S. National Science Foundation grant No. DMR-2002850 by the Simons Collaboration on Ultra-Quantum Matter which is a grant from the Simons Foundation (651440, S.S.). 
D.G.J. acknowledges support from the Leopoldina fellowship by the German National Academy of Sciences through grant no. LPDS 2020-01.
%%%%%%%%%%%%%%%%%%%%%%%%%%%%%%%%%%%%%%%%%%%%%%%%%%%%%%%%%%%%%%%%%%%%%%%%%%%%%

% Bibliography
\bibliography{ref}
\newpage


\section*{Supplementary material}

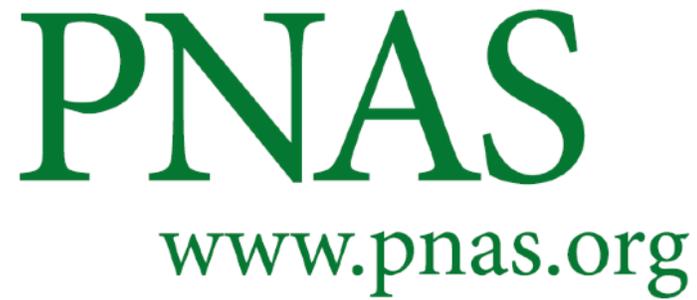

## Supplementary Information for

### Critical metallic phase in the overdoped random $t$-$J$ model

**Maine Christos, Darshan G. Joshi, Subir Sachdev and Maria Tikhanovskaya**

**Corresponding Author name.**
**E-mail: author.two@email.com**

**This PDF file includes:**

Supplementary text
Figs. S1 to S13
References for SI reference citations

## Supporting Information Text

## 1. Real frequency equations for critical metal with fermionic spinons

In this appendix we present details related to solving the saddle-point equations in Eqs. 13-18 of the main text on the real-frequency axis. We shall focus our attention at zero temperature. In addition, we also present more details of our solution and present arguments for the absence of a Fermi-liquid solution.

**A. Analytic continuation to real frequency.** We can separate the zero frequency bosonic greens function into a part which is regular at $i\omega_n = 0$ and a part which diverges at zero temperature:

$$\beta \overline{r} = \beta r - \overline{G_b}(i\omega_n = 0) \quad \text{[S1]}$$

We can then rewrite the saddle point equations in terms of $r$ and $\overline{G}_b$ as:

$$\overline{G}_b(i\omega_n = 0) - \beta r = \frac{1}{\mu_b - \Sigma_b(i\omega_n = 0)} \quad \text{[S2]}$$

$$\overline{G}_b(i\omega_n) = \frac{1}{i\omega_n + \mu_b - \Sigma_b(i\omega_n)} \quad \text{[S3]}$$

$$\Sigma_b(\tau) = -t^2 G_f(\tau) G_f(-\tau) \left[ -r + \bar{G}_b(\tau) \right] \quad \text{[S4]}$$

$$\overline{G}_f(i\omega_n) = \frac{1}{i\omega_n + \mu_f - \Sigma_f(i\omega_n)} \quad \text{[S5]}$$

$$\Sigma_f(\tau) = -J^2 G_f(\tau)^2 G_f(-\tau) + kt^2 G_f(\tau) \left[ r^2 - r(\overline{G}_b(\tau) + \overline{G}_b(-\tau)) + \overline{G}_b(\tau) \overline{G}_b(-\tau) \right] \quad \text{[S6]}$$

We wish to solve the above equations on the real frequency axis so that we may obtain physical observables like the electron and spin spectral functions. The above equations can be rewritten in terms of the spectral functions for fermions and bosons by analytically continuing the imaginary frequency greens functions to obtain the retarded greens function $G_{f/b}^R(\omega)$:

$$\rho_{f/b} = -\frac{1}{\pi} \text{Im}[G_{f/b}(i\omega_n = \omega + i0^+)] = -\frac{1}{\pi} \text{Im}[G_{f/b}^R(\omega)] \quad \text{[S7]}$$

We can similarly define the retarded self energies as $\Sigma_{f/b}^R(\omega) = \Sigma_{f/b}(i\omega_n = \omega + i0^+)$, leading to the following real frequency expressions for the saddle point equations:

$$\rho_{f/b}(\omega) = -\frac{1}{\pi} \frac{\Sigma_{f/b}^{''R}(\omega)}{(\omega - \Sigma_{f/b}^{'R}(\omega))^2 + \Sigma_{f/b}^{''R}(\omega)^2} \quad \text{[S8]}$$

Where we define the imaginary part of the retarded self energy for bosons as: $\Sigma_b^{''R}(\omega)$ as follows:

$$\Sigma_b^{''R}(\omega) = \Sigma_b^1(\omega) + \Sigma_b^2(\omega) \quad \text{[S9]}$$

with:

$$\Sigma_b^1(\omega) = t^2 r \pi \int d\omega_1 \rho_f(\omega_1) \rho_f(\omega_1 - \omega) \left[ n_f(\omega_1) n_f(\omega - \omega_1) - n_f(-\omega_1) n_f(\omega_1 - \omega) \right] \quad \text{[S10]}$$

$$\Sigma_b^2(\omega) = t^2 \pi \int d\omega_1 d\omega_2 \rho_f(\omega_1) \rho_b(\omega_2) \rho_f(\omega_1 + \omega_2 - \omega) \\ \times \left[ n_f(\omega_1) n_b(\omega_2) n_f(\omega - \omega_1 - \omega_2) + n_f(-\omega_1) n_b(-\omega_2) n_f(\omega_1 + \omega_2 - \omega) \right] \quad \text{[S11]}$$

We also define the imaginary part of the retarded self energy for fermions as $\Sigma_f^{''R}(\omega)$ as follows:

$$\Sigma_f^{''R}(\omega) = \Sigma_f^1(\omega) + \Sigma_f^2(\omega) + \Sigma_f^3(\omega) + \Sigma_f^4(\omega) \quad \text{[S12]}$$

$$\Sigma_f^1(\omega) = -J^2 \int d\omega_1 d\omega_2 \rho_f(\omega_1) \rho_f(\omega_2) \rho_f(\omega_1 + \omega_2 - \omega) \\ \times \left[ n_f(\omega_1) n_f(\omega_2) n_f(\omega - \omega_1 - \omega_2) + n_f(-\omega_1) n_f(-\omega_2) n_f(\omega_1 + \omega_2 - \omega) \right] \quad \text{[S13]}$$

$$\Sigma_f^2(\omega) = -t^2 r^2 k \pi \rho_f(\omega) \quad \text{[S14]}$$

$$\Sigma_f^3(\omega) = kt^2 r \pi \int d\omega_1 \rho_f(\omega_1) \left[ \rho_b(\omega - \omega_1) - \rho_b(\omega_1 - \omega) \right] \left[ n_b(\omega_1 - \omega) n_f(-\omega_1) - n_b(\omega - \omega_1) n_f(\omega_1) \right] \quad \text{[S15]}$$

$$\Sigma_f^4(\omega) = kt^2 \pi \int d\omega_1 d\omega_2 \rho_f(\omega_1) \rho_b(\omega_2) \rho_b(\omega_1 + \omega_2 - \omega) \\ \times \left[ n_f(\omega_1) n_b(\omega_2) n_b(\omega - \omega_1 - \omega_2) + n_f(-\omega_1) n_b(-\omega_2) n_b(\omega_1 + \omega_2 - \omega) \right] \quad \text{[S16]}$$

The constraints on the fermionic and bosonic fillings in real frequency can be rewritten as:

$$r + \int_{-\infty}^{\infty} d\omega \frac{\rho_b(\omega)}{e^{\beta\omega} - 1} = p \qquad \int_{-\infty}^{\infty} d\omega \frac{\rho_f(\omega)}{1 + e^{\beta\omega}} = \kappa - kp \qquad \text{[S17]}$$

The real parts of the bosonic and fermionic self energies are related to their imaginary self energies via the Kramers-Kronig relation. At zero temperature, the Fermi-Dirac and Bose-Einstein functions in the imaginary self energies are replaced by $\theta(-\omega)$ and $-\theta(-\omega)$ respectively. Then at zero temperature, the saddle point equations become:

$$\Sigma_b^{''R}(\omega) = \Sigma_b^1(\omega) + \Sigma_b^2(\omega) \qquad \text{[S18]}$$

$$\Sigma_b^1(\omega) = \begin{cases} -t^2 r\pi \int_0^{\omega} d\omega_1 \rho_f(\omega_1)\rho_f(\omega_1 - \omega), & \text{for } \omega > 0 \\ t^2 r\pi \int_{\omega}^{0} d\omega_1 \rho_f(\omega_1)\rho_f(\omega_1 - \omega), & \text{for } \omega < 0 \end{cases} \qquad \text{[S19]}$$

$$\Sigma_b^2(\omega) = \begin{cases} -t^2 \pi \int_0^{\omega_1+\omega_2<\omega} d\omega_1 d\omega_2 \rho_f(\omega_1)\rho_b(\omega_2)\rho_f(\omega_1 + \omega_2 - \omega), & \text{for } \omega > 0 \\ -t^2 \pi \int_{\omega_1+\omega_2>\omega}^{0} d\omega_1 d\omega_2 \rho_f(\omega_1)\rho_b(\omega_2)\rho_f(\omega_1 + \omega_2 - \omega), & \text{for } \omega < 0 \end{cases} \qquad \text{[S20]}$$

$$\Sigma_f^{''R}(\omega) = \Sigma_f^1(\omega) + \Sigma_f^2(\omega) + \Sigma_f^3(\omega) + \Sigma_f^4(\omega) \qquad \text{[S21]}$$

$$\Sigma_f^1(\omega) = \begin{cases} -J^2 \pi \int_0^{\omega_1+\omega_2<\omega} d\omega_1 d\omega_2 \rho_f(\omega_1)\rho_f(\omega_2)\rho_f(\omega_1 + \omega_2 - \omega), & \text{for } \omega > 0 \\ -J^2 \pi \int_{\omega_1+\omega_2>\omega}^{0} d\omega_1 d\omega_2 \rho_f(\omega_1)\rho_f(\omega_2)\rho_f(\omega_1 + \omega_2 - \omega), & \text{for } \omega < 0 \end{cases} \qquad \text{[S22]}$$

$$\Sigma_f^2(\omega) = -t^2 r^2 k\pi \rho_f(\omega) \qquad \text{[S23]}$$

$$\Sigma_f^3(\omega) = \begin{cases} -kt^2 r\pi \int_0^{\omega} d\omega_1 \rho_f(\omega_1)[\rho_b(\omega - \omega_1) - \rho_b(\omega_1 - \omega)], & \text{for } \omega > 0 \\ kt^2 r\pi \int_{\omega}^{0} d\omega_1 \rho_f(\omega_1)[\rho_b(\omega - \omega_1) - \rho_b(\omega_1 - \omega)], & \text{for } \omega < 0 \end{cases} \qquad \text{[S24]}$$

$$\Sigma_f^4(\omega) = \begin{cases} kt^2 \pi \int_0^{\omega_1+\omega_2<\omega} d\omega_1 d\omega_2 \rho_f(\omega_1)\rho_b(\omega_2)\rho_b(\omega_1 + \omega_2 - \omega), & \text{for } \omega > 0 \\ kt^2 \pi \int_{\omega_1+\omega_2>\omega}^{0} d\omega_1 d\omega_2 \rho_f(\omega_1)\rho_b(\omega_2)\rho_b(\omega_1 + \omega_2 - \omega), & \text{for } \omega < 0 \end{cases} \qquad \text{[S25]}$$

Subject to the filling constraints:

$$r - \int_{-\infty}^{0} d\omega \rho_b(\omega) = p \qquad \int_{-\infty}^{0} d\omega \rho_f(\omega) = \kappa - kp \qquad \text{[S26]}$$

**B. Absence of a Fermi-liquid solution.** In the above saddle point equations, the low frequency structure for $r > 0$ implies $\overline{G}_b(i\omega_n = 0) \sim 1/|\omega|$. Such low frequency behavior leads to a logarithmically diverging boson density and $r$, meaning that we must have $r = 0$ at $T = 0$. We then set $r = 0$ and begin looking for critical solutions to the above equations at $T = 0$ of the following form:

$$G_a(i\omega_n) = -iC_a \begin{pmatrix} e^{-i\theta_a} \\ -e^{i\theta_a} \end{pmatrix} \frac{1}{\omega^{1-2\Delta_a}} \qquad \text{[S27]}$$

and corresponding low frequency form for the spectral density:

$$\rho_a(\omega) = \frac{C_a}{\pi} \begin{pmatrix} \sin(\pi\Delta_a + \theta_{f/b}) \\ \sin(\pi\Delta_a - \theta_a) \end{pmatrix} \frac{1}{\omega^{1-2\Delta_a}} \qquad \text{[S28]}$$

where $a = f/b$ and with the additional constraint $\Delta_f + \Delta_b = \frac{1}{2}$. To resolve the low frequency divergences in the spectral densities in our numerical integration, we absorb the the low frequency $\omega$ dependence of the spectral functions into the numerical integration measure such that the numerical integrals do not contain any divergent pieces. We use $4 \times 10^4$ frequency points on our real $\omega$ axis and a frequency spacing which is smaller at small $\omega$ and larger at large $\omega$.

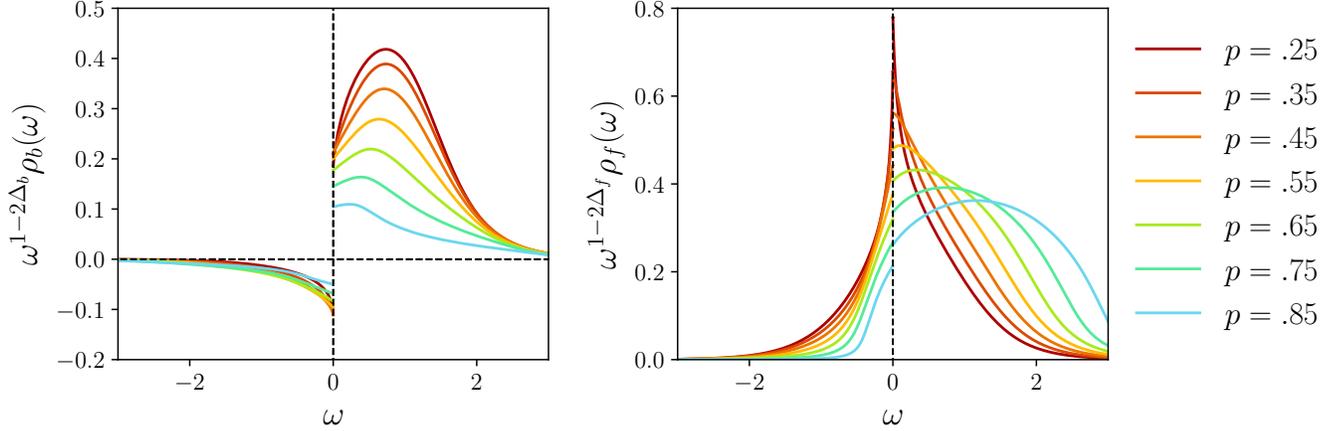

**Fig. S1.** Plot of the boson spectral density (left) and fermion spectral density (right) obtained at $t = J = 1$. Both the boson and fermion spectral densities show deviations from the conformal power law behavior for small frequencies.

**C. Fixing the low energy solution.** In this section we will detail how we fix the coefficients $C_f$, $C_b$, $\Delta_f$, $\Delta_b$, $\theta_f$, and $\theta_b$ in Eq. S28. The filling constraints for bosons and fermions will be imposed by the initial conditions we select for $\theta_f$, $\theta_b$, $\Delta_f$, and $\Delta_b$. In constraining these coefficients in the low energy solution, we first solve for $\theta_f$ and $\theta_b$ as a function of $\Delta_b$ at fixed doping. We do this for a given $\Delta_b$ and $p$ by separately solving the two Luttinger constraints for fermions and bosons to obtain corresponding values for $\theta_f$ and $\theta_b$,

$$\frac{\theta_f}{\pi} + \left(\frac{1}{2} - \Delta_f\right) \frac{\sin(2\theta_f)}{\sin(2\pi\Delta_f)} = \frac{1}{2} - \kappa + kp,$$
$$\frac{\theta_b}{\pi} + \left(\frac{1}{2} - \Delta_b\right) \frac{\sin(2\theta_b)}{\sin(2\pi\Delta_b)} = \frac{1}{2} + p. \quad \text{[S29]}$$

For a fixed doping, we can then use $\theta_f$ and $\theta_b$ calculated as a function of $\Delta_b$ to compute the coefficient $k$ as a function of $\Delta_b$ using its expression in terms of the self-consistency conditions,

$$k = -\frac{\Gamma(2-2\Delta_f)\Gamma(2\Delta_f)\sin(\pi\Delta_f + \theta_f)\sin(\pi\Delta_f - \theta_f)}{\Gamma(2-2\Delta_b)\Gamma(2\Delta_b)\sin(\pi\Delta_b + \theta_b)\sin(\pi\Delta_b - \theta_b)}, \quad \text{[S30]}$$

We then fix $\Delta_b$ by selecting the value of $\Delta_b$ at which $k$ attains its physical value $k = \frac{1}{2}$. We show for 3 different dopings the $k$ we compute in this way as a function of $\Delta_b$ in Fig. S2 and the values of $\theta_f$, $\theta_b$, $\Delta_f$, $\Delta_b$ we obtain by choosing $k = \frac{1}{2}$ in Fig. S3. We then have fixed every free parameter in Eq. S28 except for $C_b$ and $C_f$. $C_f$ and $C_b$ are parameters which are unconstrained in the original low frequency ansatz and generally have values which depend on the full $\omega$ dependent solutions $\rho_f(\omega)$ and $\rho_b(\omega)$. We solve for these constants at each step of our iterative procedure by taking the Hilbert transform of the spectral densities to obtain the bosonic and fermions greens functions from which we may extract the constant $C_f$ at low frequency and a small temperature $T = 10^{-5}$. $C_b$ can then be obtained from $C_f$ via the self consistency equations and these extracted values for $C_f$ and $C_b$ are then plugged in to the next step of numerical iteration. Their values are shown in Fig. S3. Results for the boson and fermion spectral densities computed in this manner are shown in Fig. S1.

We also compute the physical observables, the spin and electron spectral functions, $\rho_s(\omega) = \frac{1}{\pi}\chi''(\omega)$ and $\rho_c(\omega)$. In the limit $r = 0$, these observables have the following expressions on the real frequency axis:

$$\rho_c(\omega) = \int_{-\infty}^{\infty} \rho_f(\omega_1)\rho_b(\omega_1 - \omega) \left[n_f(\omega_1) + n_b(\omega_1 - \omega)\right], \quad \text{[S31]}$$

$$\chi''(\omega) = \pi\rho_s(\omega) = \pi \int_{-\infty}^{\infty} \rho_f(\omega_1)\rho_f(\omega_1 - \omega) \left[n_f(\omega_1 - \omega) - n_f(\omega_1)\right]. \quad \text{[S32]}$$

Results for these spectral functions appeared in Fig. 3. The electron spectral density satisfies the sum rule,

$$\int_{-\infty}^{\infty} d\omega \, \rho_c(\omega) = \frac{1+p}{2}. \quad \text{[S33]}$$

Note that for canonical fermions this sum rule results in unity. However, here the electron operator is being fractionalized and the Hilbert-space has only three states. Hence the sum rule is modified. There is a also a simple way to understand this modification. The electron spectral density sum is equal to $\langle\{c_\alpha, c_\alpha^\dagger\}\rangle$. For canonical fermions, $\{c_\alpha, c_\alpha^\dagger\} = 1$, but in our case

Maine Christos, Darshan G. Joshi, Subir Sachdev and Maria Tikhanovskaya

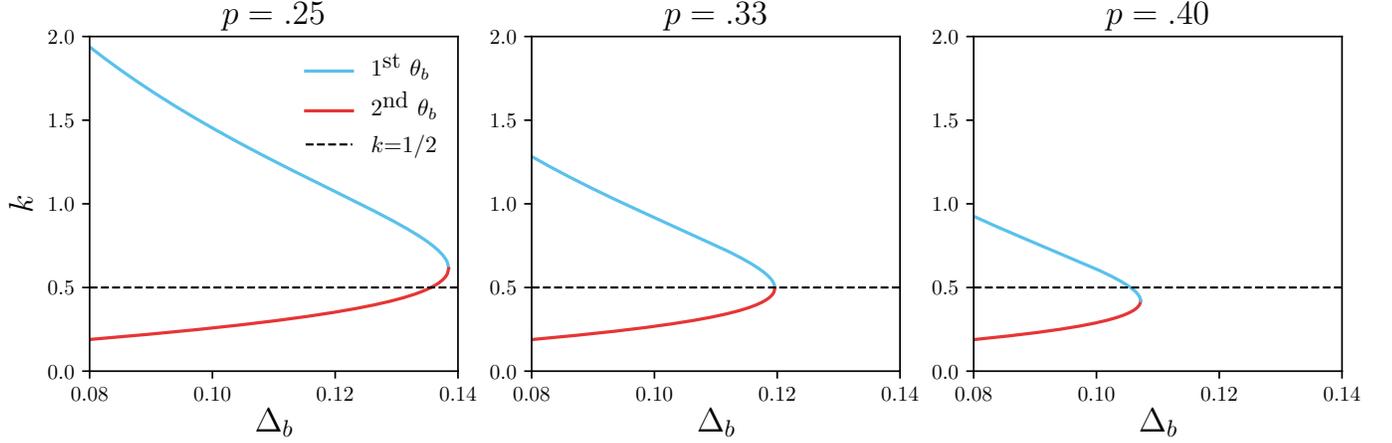

**Fig. S2.** Plot of the value of $k$ as computed from the zero frequency self consistency conditions for three different dopings. The values of $\theta_f$ and $\theta_b$ are computed from the Luttinger relations at fixed $\Delta_b$ (and fixed $\Delta_f = \frac{1}{2} - \Delta_b$) and are then used to compute $k$ via the self consistency condition. We find in general there are two possible values of $\theta_b$ which are compatible with the Luttinger constraints, leading to the two distinct branches of the function for $k$ shown in red and blue above. The correct value of $\Delta_b$ is selected by choosing the value of $\Delta_b$ where $k = \frac{1}{2}$. The physical value $k = \frac{1}{2}$ is denoted with a dashed black line. As is shown in the above plot, for $p > \frac{1}{3}$, this leads to selecting the first (smallest) value of $\theta_b$ and for $p < \frac{1}{3}$ leads to selecting the second (largest) value of $\theta_b$, with the two values of $\theta_b$ coinciding at $p = \frac{1}{3}$ and $k = \frac{1}{2}$.

$\{c_\alpha, c_\alpha^\dagger\} = b^\dagger b + f_\alpha^\dagger f_\alpha$, which gives the modified sum rule in Eq. S33. The sum rule for the spin spectral density is not easy to obtain at finite temperature. But at zero temperature we have,

$$\int_0^\infty \rho_s(\omega) = \frac{1-p^2}{4} \, . \tag{S34}$$

**D. Varying $J$.** In this section, we will discuss the behavior of the spectral functions as $J$ is varied between $0 \leq J \leq t$. For the case where $0 < \Delta_b < \frac{1}{4}$ and $\frac{1}{4} < \Delta_f < \frac{1}{2}$, the Luttinger constraints and saddle point equations exactly at zero frequency have no dependence on $J$, meaning the values of $\Delta_b$, $\Delta_f$, $\theta_b$, and $\theta_f$ we solve for do not depend on the value of $J$ (though $C_f$ and $C_b$ which depend on the full $\omega$ dependent forms of $\rho_f(\omega)$ and $\rho_b(\omega)$ do depend on $J$). We then follow the same procedure as for $t = J = 1$ and solve the saddle point equations for $0 \geq J \geq t = 1$. We find mostly qualitative changes for high doping, however we find a curious feature in the electron spectral density at low doping as shown in Fig. S4. We find that for small enough doping around $p < .3$ that the electron spectral density acquires a second local maximum or peak at finite $\omega < 0$ for small $J$ which does not appear for larger values of $J$ or larger values of doping $p$. We find that this peak originates from a similar peak in the bosonic spectral density that occurs when $\omega = \Sigma_b'(\omega)$ and $\rho_b(\omega) = -\frac{1}{\pi \Sigma_b''(\omega)}$. While we do not have an analytic reason for this behavior of the boson and electron spectral functions, we reason that this feature originates from strong corrections to the conformal solutions for $\rho_b$ and $\rho_c$ away from $\omega = 0$.

## 2. Additional results from imaginary-frequency calculation

In this appendix we present additional results obtained from the imaginary-frequency analysis. In particular, we will discuss an independent way to extract the exponent $\Delta_f$ and numerical analytic continuation to real frequencies.

**A. Holon and spinon Green's function in the critical metal phase.** The strategy to solve the saddle-point equations, Eqs. 13-18, is straightforward. We first fix $\bar{r}$ and solve Eqs. 13-16 until we find a converged solution. Then we check if these solutions give the same doping value using Eqs. 17 and 18. If it does then we have our final solution for a given doping. Else, we go back to step one and change $\bar{r}$, and repeat the next steps until we find the final solution. In this fashion, we obtain the holon and spinon Green's functions on the imaginary frequency axis, which are shown in Fig. S5. Recall that in the critical metal phase, holon is a bosonic and spinons are fermionic operators. We have checked that these satisfy the respective sum rules: $G_a(0^+) - \zeta_a G_a(\beta^-) = -1 = -\int_{-\infty}^\infty \rho_a(\omega) d\omega$, where $\zeta_a = \pm 1$ for $a = b, f$ respectively. Once we have the holon and spinon Green's functions it is then straightforward to obtain the electron Green's function and the spin correlator.

As discussed in the main text, using the temperature dependence of the holon Green's function at $i\omega = 0$ we can estimate the exponent, $\Delta_b$. In Fig. S6 we show the data points used to estimated $\Delta_b$ at different dopings.

**B. Spin-correlator in the critical metal phase.** Let us consider the spin correlator,

$$\frac{\langle \vec{S}(\tau) \cdot \vec{S}(0) \rangle}{2} = Q(\tau) = G_f(\tau) G_f(-\tau) \, . \tag{S35}$$

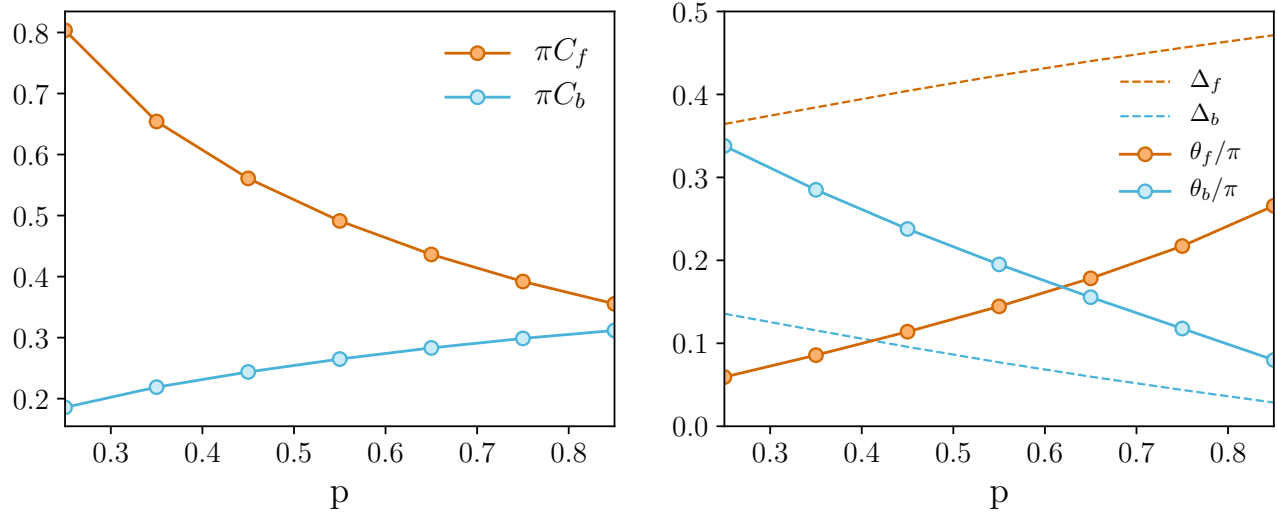

**Fig. S3.** In the left plot, we show the converged values of $C_f$ and $C_b$ as a function of doping for $t = J = 1$. In the right plot, we show the doping dependence of $\Delta_b$, $\Delta_f$, $\theta_b$, and $\theta_f$, obtained by simultaneously solving the two Luttinger relations and the two constraints, as discussed in the text.

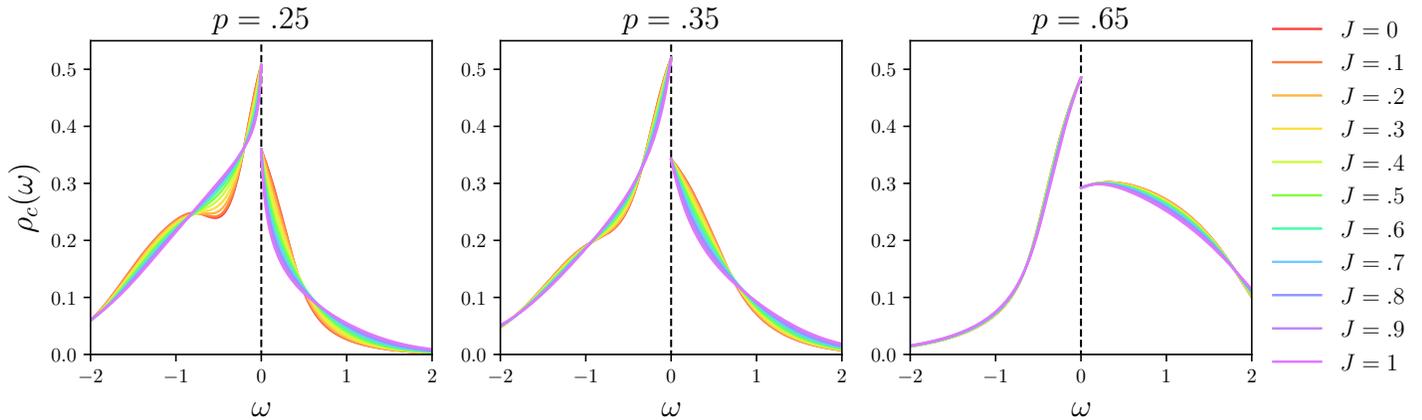

**Fig. S4.** Plot showing how the electron spectral density varies with $J$ for three distinct dopings and $t = 1$.

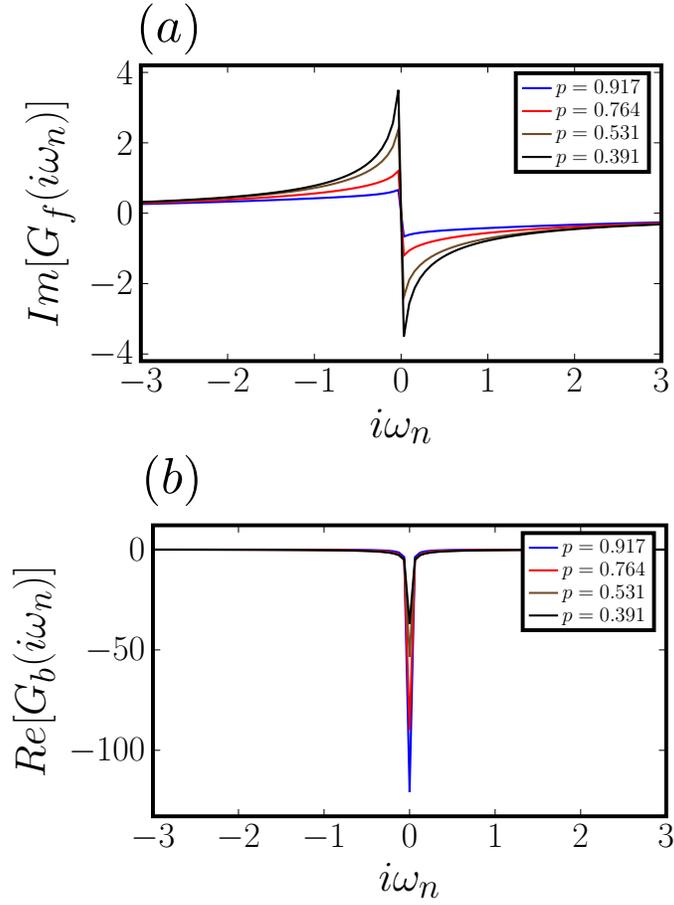

**Fig. S5.** (a) Imaginary part of the fermionic spinon Green's function on the Matsubara-frequency axis for different dopings at $T = 0.01$. (b) Real part of the bosonic holon Green's function on the Matsubara-frequency axis for different dopings at $T = 0.01$. We have used $t = J = 1$.

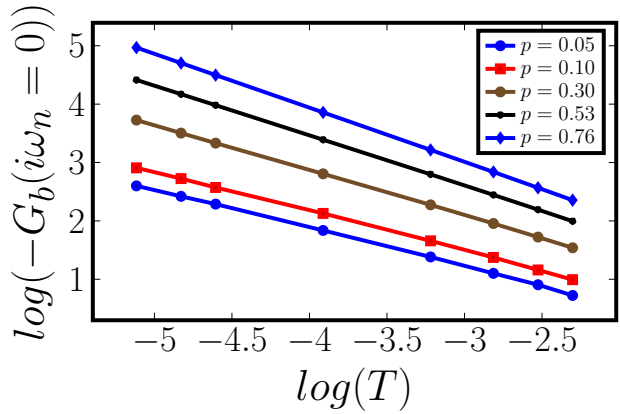

**Fig. S6.** Log-log plot of the relation in Eq. 26 for different dopings. We find that the data fits a linear curve to a good accuracy. Thus we can use the value of slopes of these curves to determine $\Delta_b$, as discussed in Section B.

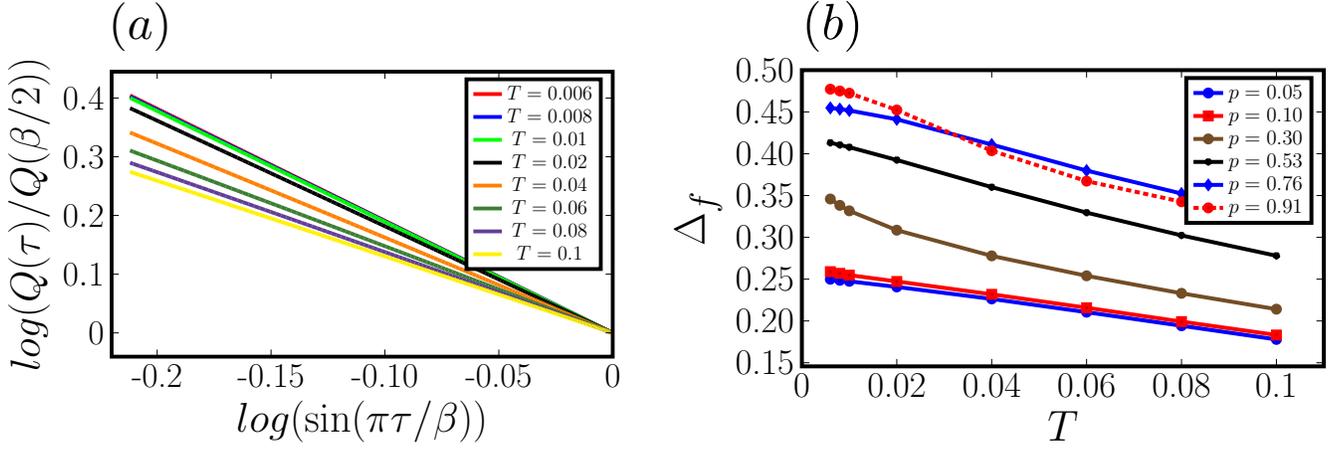

**Fig. S7.** (a) Log-log plot of Eq. S36 near $\tau = \beta/2$ for different temperatures at $p = 0.91$. The data considered for the plot is in the range $1/2 \leq \tau/\beta \lesssim 0.7$. (b) Plot of $\Delta_f$ as a function of temperature $T$ for different dopings. Note that $4\Delta_f$ is the exponent of the spin correlation.

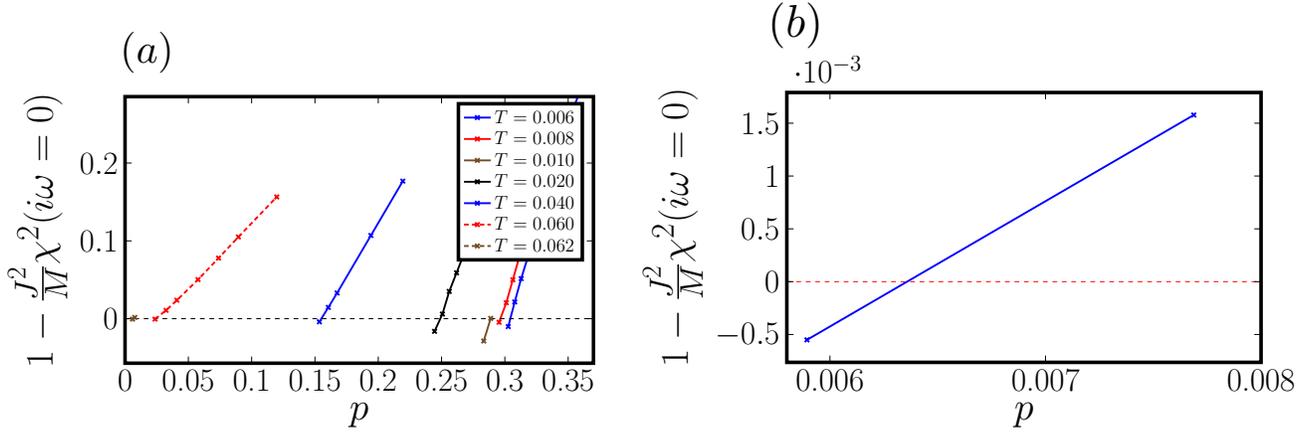

**Fig. S8.** (a) Coefficient of quadratic term in Eq. 27 as a function of doping for different temperature. As explained in the text, a sign change of this coefficient gives the critical doping. (b) Enlarged plot for T=0.062.

In the critical metal phase, the solution has a conformal form in the low-frequency or large-time limit. At a finite temperature, in terms of the imaginary time this limit corresponds to $\tau = \beta/2$, where we expect the conformal form,

$$\frac{Q(\tau)}{Q(\beta/2)} \sim \frac{1}{[\sin(\pi\tau/\beta)]^{4\Delta_f}}. \qquad [\text{S36}]$$

Thus if we make a log-log plot then we can extract $\Delta_f$ from the slope of the resulting linear curve. In Fig. S7A we show such a plot for several different temperatures at a fixed doping, $p = 0.91$. These curves fit to a linear curve. In Fig. S7 (b) we plot the extracted value of $\Delta_f$ as a function of temperature for different dopings. Remarkably, the extrapolated zero-temperature values of $\Delta_f$ in the overdoping region for $p \gtrsim 0.3$ are in agreement with those found analytically earlier. At lower dopings we find some disagreement with the analytic result.

As discussed in Section C, using the local spin susceptibility, $\chi(i\omega = 0)$, we can estimate the critical doping below which the spin-glass is stabilized. We therefore plot the coefficient of quadratic term in Eq. 27 as a function of doping for different temperatures in Fig. S8. This gives the critical doping as a function of temperature, which is used to plot the phase diagram in Fig. 1B.

**C. Numerical analytic continuation.** We also perform numerical analytic continuation to real frequency. In general, performing analytic continuation is an ill-posed problem if the function on the imaginary axis is known only at a finite number of points. There are several techniques to do analytic continuation. But for simplicity we use the Pade approximation method. This technique parametrizes the function on imaginary axis as a ratio of two polynomials or by terminating a continued fraction. There are several ways for implementing Pade approximation. We adopt the simple strategy outlined in Ref. (1) of evaluating the coefficients of the two polynomials recursively, which is based on Thiele's reciprocal difference method. Details of the algorithm can be found in the Appendix of Ref. (1). Briefly, we first solve the saddle-point equations on the imaginary-frequency axis to obtain the required Green's function, say $F(i\omega)$, at non-negative Matsubara frequencies. The number of Matsubara

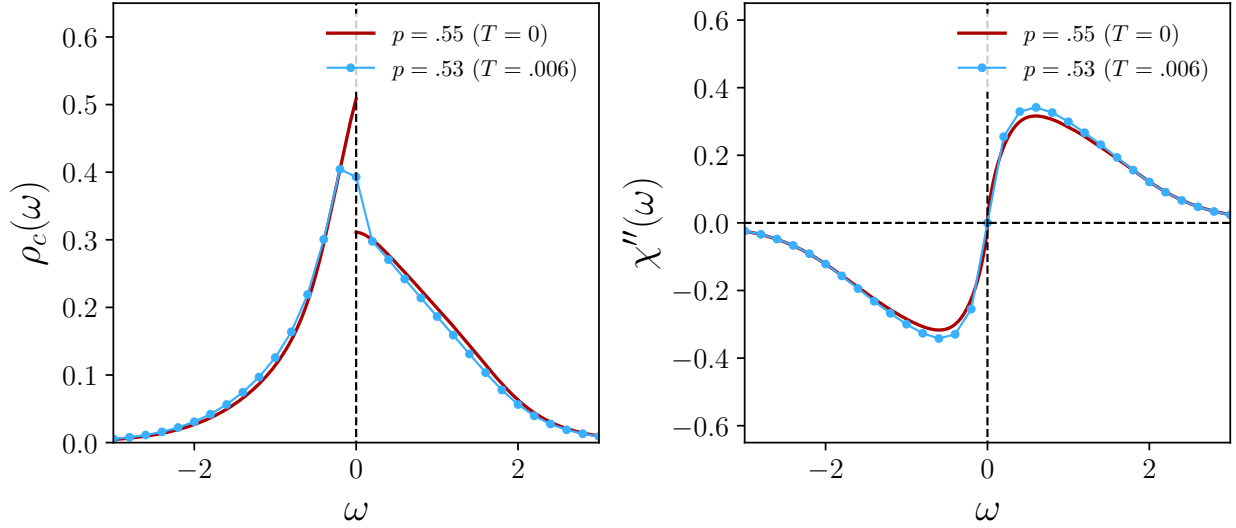

**Fig. S9.** Comparison of electron and spin spectral densities obtained for similar dopings from the $T=0$ real-frequency numerics and analytic continuation of the imaginary frequency solution to real frequency at $T=0.006$.

frequencies used in our calculation is $10^5$. Then we evaluate the required polynomials, $A_n(z)$ and $B_n(z)$, to approximate the imaginary-frequency function, $F(z) = A_n(z)/B_n(z)$. The accuracy of these polynomials depends on the number of Pade points, $n$, and in our calculation we find that $n = 200$ points are sufficient to obtain accurate results. We have checked our results by increasing or decreasing $n$ and it does not result in any significant improvement. The resulting ratio of polynomials then corresponds to the retarded Green's function on real-frequency axis, once we identify $z = \omega + i\eta$. Using the imaginary part of this function we then readily obtain the spectral densities.

In the critical non-Fermi-liquid metal phase, we find the electron and spin spectral densities using this approach. In Fig. S9, we plot the results of analytic continuation at $T = 0.006$ alongside the results from real-frequency analysis at zero temperature. The agreement is remarkably good. We have also checked that the electron and spin spectral densities obtained via numerical analytic continuation satisfy to a good accuracy the appropriate sum rules in Eqs. S33 and S34. In this phase, as shown earlier, the spinon and holon spectral densities have a $\omega^{1-2\Delta_a}$ divergence. In principle, we can adopt our method if we scale this divergence, as done in the real-frequency analysis. However, we do not attempt to do it here. Anyways, these quantities are not gauge-invariant observables.

In the spin-glass phase the holon and fermion spectral densities do not have divergences. So in this case we calculate the holon and spinon spectral densities in addition to the spin and electron spectral densities. In Figs. S10 and S11 we plot these spectral densities at $p = 0.027$ and $p = 0.2$ respectively, obtained by analytic continuation of imaginary-frequency results at $T = 0.01$. Alongside it we also plot the results obtained at $T = 0$ for a nearby doping value. The agreement is remarkable. The spectral densities obtained via numerical analytic continuation satisfy the respective sum rules listed in Eqs. S82, S85, and S86.

## 3. Insulating SY model with bosonic spinons

This appendix will recall the solution of an insulating spin glass obtained using bosonic spinons in Ref. (2). We consider the spin model

$$H = \frac{1}{\sqrt{NM}} \sum_{i<j=1}^{N} \sum_{\alpha,\beta=1}^{M} J_{ij} \mathcal{S}_\beta^\alpha(i) \mathcal{S}_\alpha^\beta(j) \quad [S37]$$

where $\mathcal{S}_\beta^\alpha(i) = [\mathcal{S}_\alpha^\beta(i)]^\dagger$ are generators of SU($M$) on each site $i$. Each site contains states corresponding to the symmetric product of $\kappa M$ fundamentals, and these are realized by bosonic spinons with

$$\mathcal{S}_\beta^\alpha(i) = \mathfrak{b}_\beta^\dagger(i)\mathfrak{b}^\alpha(i) - \kappa \delta_\beta^\alpha, \quad \sum_\alpha \mathfrak{b}_\alpha^\dagger(i)\mathfrak{b}^\alpha(i) = \kappa M \quad [S38]$$

on each site $i$. We have made the spin operators traceless. The exchange constants $J_{ij}$ are mutually uncorrelated and selected with probability $P(J_{ij}) \sim \exp(-J_{ij}^2/(2J^2))$.

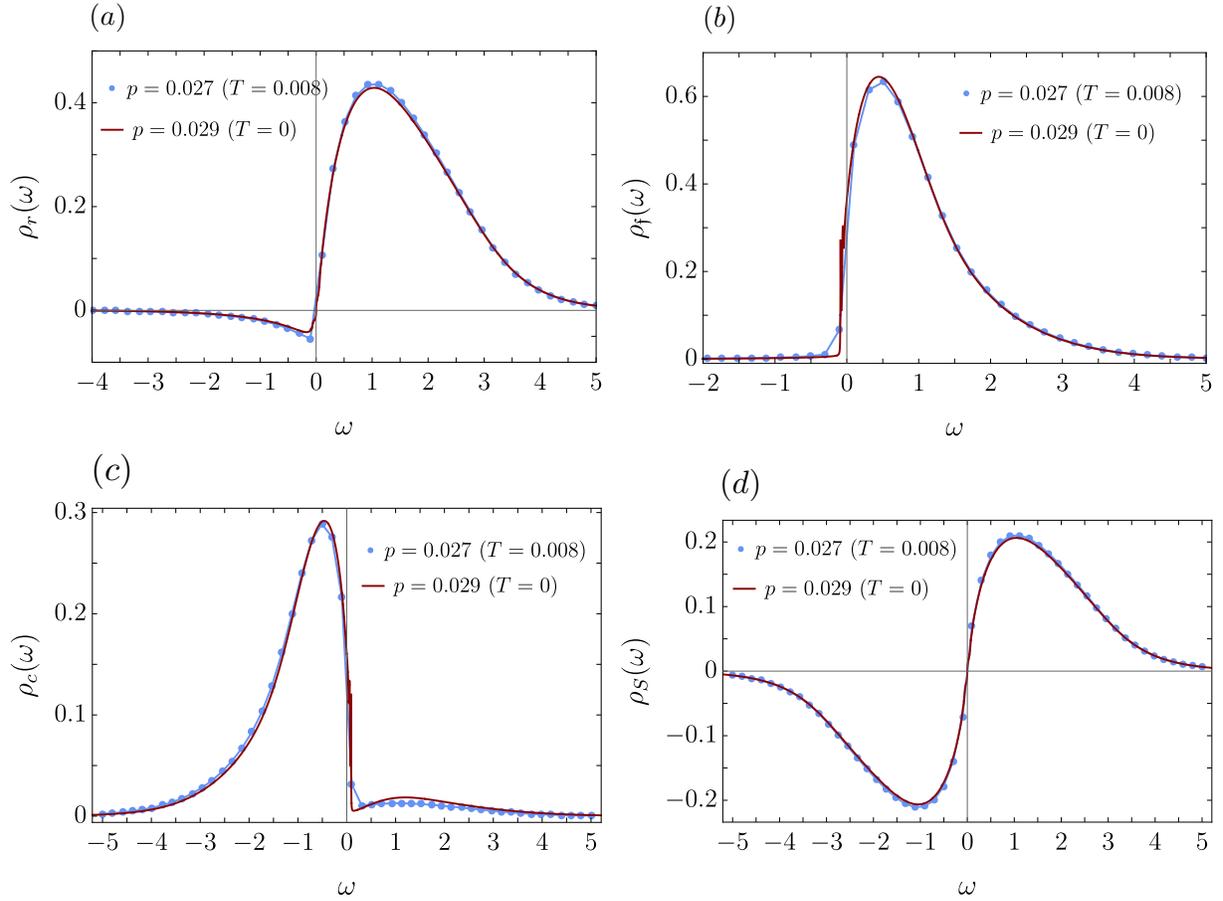

**Fig. S10.** Comparison of spectral densities in the spin glass phase at small doping obtained via analytic continuation at $T = 0.008$ for $p = 0.027$ and via real frequency analysis at zero temperature for $p = 0.029$. From (a) to (d), we plot the holon, spinon, electron, and spin spectral densities respectively.

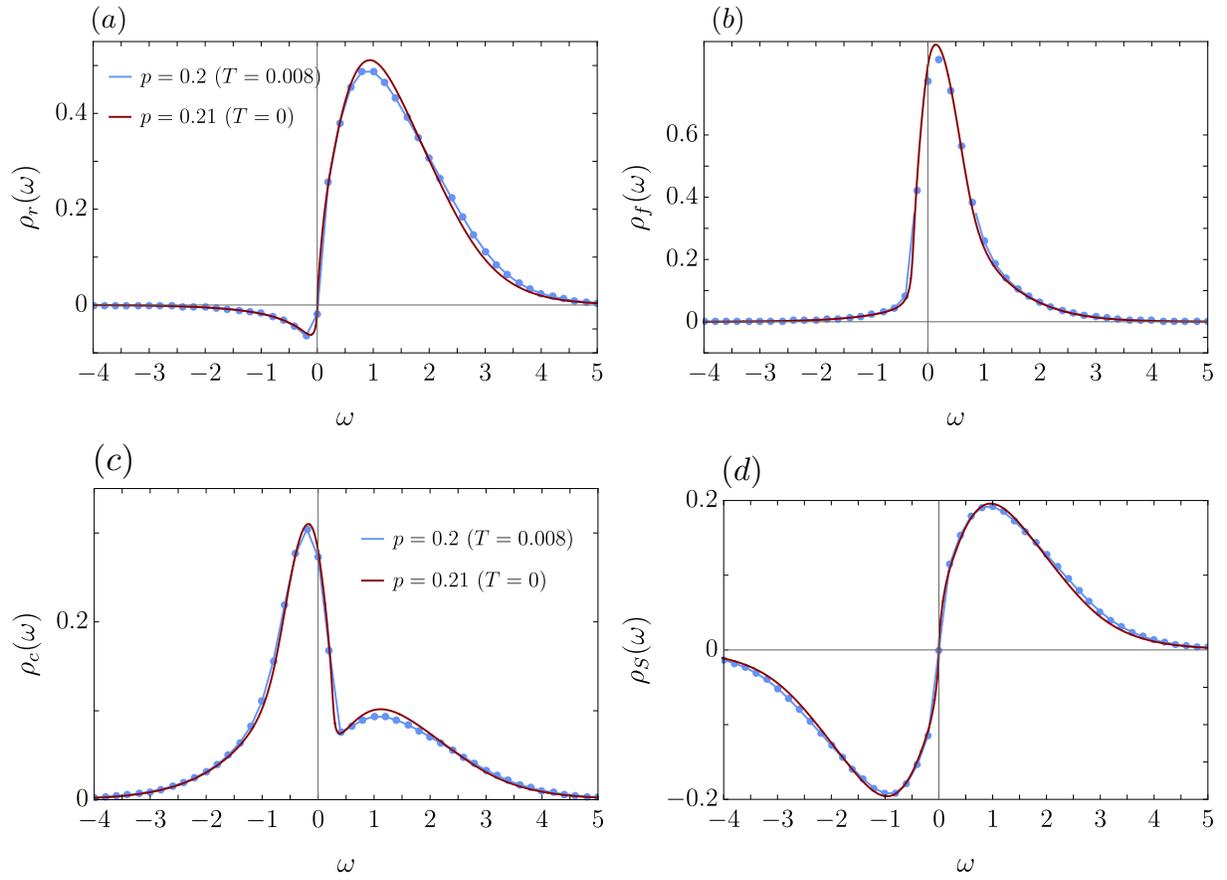

**Fig. S11.** Comparison of spectral densities in the spin glass phase at small doping obtained via analytic continuation at $T = 0.008$ for $p = 0.2$ and via real frequency analysis at zero temperature for $p = 0.21$. From (a) to (d), we plot the holon, spinon, electron, and spin spectral densities respectively.

We introduce replicas $a = 1 \ldots n$, and average over $J_{ij}$ to obtain the averaged, replicated partition function

$$\overline{\mathcal{Z}^n} = \int \mathcal{D}\mathfrak{b}_a^\alpha(i,\tau)\mathcal{D}\lambda_a(i,\tau) \exp\left[-S_B - S_J\right]$$

$$S_B = \sum_i \int d\tau \left[\mathfrak{b}_{a\alpha}^\dagger(i)\partial_\tau \mathfrak{b}_a^\alpha(i) + i\lambda_a(i)\left(\mathfrak{b}_{a\alpha}^\dagger(i)\mathfrak{b}_a^\alpha(i) - \kappa M\right)\right]$$

$$S_J = -\frac{J^2}{4NM} \int d\tau d\tau' \left[\sum_i \mathcal{S}_{a\beta}^\alpha(i,\tau)\mathcal{S}_{b\delta}^\gamma(i,\tau')\right]\left[\sum_j \mathcal{S}_{a\alpha}^\beta(j,\tau)\mathcal{S}_{b\gamma}^\delta(j,\tau')\right] \quad [S39]$$

We can now decouple $S_J$ with a Hubbard-Stratonovich field $Q_{ab,\beta\delta}^{\alpha\gamma}(\tau,\tau')$ and take the large $N$ limit. Then the problem reduced to finding saddle points of the action

$$\frac{\mathcal{S}[Q]}{N} = \frac{J^2}{4M} \int d\tau d\tau' |Q_{ab,\beta\delta}^{\alpha\gamma}(\tau,\tau')|^2 - \ln \mathcal{Z}_f[Q] \quad [S40]$$

where $\mathcal{Z}_f[Q]$ is the single site partition function

$$\mathcal{Z}_b[Q] = \int \mathcal{D}b_a^\alpha(\tau)\mathcal{D}\lambda_a(\tau) \exp\left[-S_B - S_b\right] \quad [S41]$$

$$S_B = \int d\tau \left[\mathfrak{b}_{a\alpha}^\dagger \partial_\tau \mathfrak{b}_a^\alpha + i\lambda_a \left(\mathfrak{b}_{a\alpha}^\dagger \mathfrak{b}_a^\alpha - \kappa M\right)\right] \quad [S42]$$

$$S_b = -\frac{J^2}{2M} \int d\tau d\tau' Q_{ab,\beta\delta}^{\alpha\gamma}(\tau,\tau') \left[\mathfrak{b}_{a\alpha}^\dagger(\tau)\mathfrak{b}_a^\beta(\tau) - \kappa\delta_\alpha^\beta\right]\left[\mathfrak{b}_{b\gamma}^\dagger(\tau')\mathfrak{b}_b^\delta(\tau') - \kappa\delta_\gamma^\delta\right] \quad [S43]$$

Note that we have taken the $N \to \infty$ limit, and there is no remaining path integral over $Q$. We simply have to find the saddle points of $\mathcal{S}[Q]$ in Eq. S40. Let us assume that the saddle point does not break spin rotation symmetry: this is true in both the spin glass, and spin liquid phases. So we take the ansatz

$$Q_{ab,\beta\delta}^{\alpha\gamma}(\tau,\tau') = \delta_\delta^\alpha \delta_\beta^\gamma Q_{ab}(\tau-\tau') \quad [S44]$$

where $Q_{ab}(\tau)$ is a real function. Also, because there is no path integral over $Q$, we can also assume from now on that $Q_{ab}(\tau)$ is independent of $\tau$ for $a \neq b$. Then Eq. S40 is replaced by

$$\frac{\mathcal{S}[Q]}{N} = \frac{J^2 M}{4} \int d\tau d\tau' [Q_{ab}(\tau-\tau')]^2 - \ln \mathcal{Z}_b[Q] \quad [S45]$$

while Eq. S43 is replaced by

$$S_b = -\frac{J^2}{2M} \int d\tau d\tau' Q_{ab}(\tau-\tau') \left[\mathfrak{b}_{a\alpha}^\dagger(\tau)\mathfrak{b}_a^\beta(\tau)\mathfrak{b}_{b\beta}^\dagger(\tau')\mathfrak{b}_b^\alpha(\tau') - \kappa^2 M\right] \quad [S46]$$

The saddle point equations for $Q$ from Eqs. S45 and S46 are

$$Q_{ab}(\tau-\tau') = \frac{1}{M^2} \left\langle \mathfrak{b}_{a\alpha}^\dagger(\tau)\mathfrak{b}_a^\beta(\tau)\mathfrak{b}_{b\beta}^\dagger(\tau')\mathfrak{b}_b^\alpha(\tau') \right\rangle_{\mathcal{Z}_b[Q]} - \frac{\kappa^2}{M} \quad [S47]$$

Now we need to evaluate $\mathcal{Z}_f[Q]$. This is conveniently done using $G$-$\Sigma$ theory, where we define

$$G_{ab}(\tau,\tau') = -\frac{1}{M}\sum_\alpha \mathfrak{b}_a^\alpha(\tau)\mathfrak{b}_{b\alpha}^\dagger(\tau') \quad [S48]$$

Then we can write

$$\mathcal{Z}_b[Q] = \int \mathcal{D}G_{ab}(\tau,\tau')\Sigma_{ab}(\tau,\tau')\mathcal{D}\lambda_a(\tau) \exp\left[-MI[Q] - \frac{\kappa^2 J^2}{2}\int d\tau d\tau' \sum_{a,b} Q_{ab}(\tau-\tau')\right]$$

$$I[Q] = \ln\det\left[-\delta'(\tau-\tau')\delta_{ab} - i\lambda_a(\tau)\delta(\tau-\tau')\delta_{ab} - \Sigma_{ab}(\tau,\tau')\right] - i\kappa\int d\tau \lambda_a(\tau)$$

$$+ \int d\tau d\tau' \left[\Sigma_{ab}(\tau,\tau')G_{ba}(\tau',\tau) - \frac{J^2}{2}Q_{ab}(\tau-\tau')G_{ab}(\tau,\tau')G_{ba}(\tau',\tau)\right] \quad [S49]$$

Unlike the fermionic case, it is sufficient to work at $M = \infty$ to obtain the spin glass phase, and there is no need to consider $1/M$ corrections. The large $M$ saddle-point equations of Eq. S49 are

$$\Sigma_{ab}(\tau) = J^2 Q_{ab}(\tau) G_{ab}(\tau) \quad [S50]$$

$$G_{ab}(i\omega) = \left[(i\omega + \overline{\lambda})\delta_{ab} - \Sigma_{ab}(i\omega)\right]^{-1} \quad [S51]$$

where $i\lambda = -\overline{\lambda}$ at the saddle-point. The saddle-point equation Eq. S47 now becomes

$$Q_{ab}(\tau) = G_{ab}(\tau)G_{ba}(-\tau) \quad [\text{S52}]$$

Then Eqs. S50, S51 and S52 combine to yield the bosonic SY equations (3) for replica diagonal case. We also have to include the saddle-point equation for $\overline{\lambda}$ which is

$$T \sum_{\omega_n} G_{aa}(i\omega_n) = -\kappa \quad [\text{S53}]$$

**A. Spin glass.** The spin glass phase was described in Ref. (2). In this phase, we must include replica off-diagonal components for $G_{ab}(\omega_n = 0)$, while other $\omega_n$ remain replica diagonal. We characterize the replica off-diagonal components of $G_{ab}(\tau)$ by a $n \times n$ matrix $g_{ab}$ which is independent of $\tau$. In other words, in frequency space

$$G_{ab}(i\omega_n) = -\beta g_{ab}\delta_{\omega_n,0}, \quad a \neq b \quad [\text{S54}]$$

In Parisi's theory of the infinite-range spin glass, the $n \to 0$ limit of the matrix $g_{ab}$ is defined by a function $g(u)$, with $u \in [0,1]$. We make the one-step replica symmetry breaking ansatz

$$\begin{aligned} g(u) &= g, \quad \text{for } x < u < 1 \\ &= 0, \quad \text{for } u < x. \end{aligned} \quad [\text{S55}]$$

*i.e.* it is a step function at the 'breakpoint' $u = x$. The matrix $g_{ab}$ is therefore fully defined by the values of $g$ and $x$. For the replica diagonal components, it is convenient to define

$$G_{aa}(i\omega_n) = -\beta g \delta_{\omega_n,0} + G_r(\omega_n) \quad [\text{S56}]$$

At the moment, there is no prescribed frequency dependence for $G_r$, and so this ansatz can be made without loss of generality. However, as we will see below, this ansatz is convenient because it leads to solutions in which $G_r$ is a smooth function of frequency.

Similarly, for the self-energy, we have from Eqs. S50 and S52

$$\Sigma_{ab}(i\omega_n) = -\beta J^2 g_{ab}^3 \delta_{\omega_n,0}, \quad a \neq b, \quad [\text{S57}]$$

and for the diagonal component we write

$$\Sigma_{aa}(i\omega_n) = -\beta J^2 g^3 \delta_{\omega_n,0} + \Sigma_r(\omega_n) \quad [\text{S58}]$$

Now we need to insert the ansatz Eqs. S54, S55, S56, S57, S58 into Eqs. S50, S51, S52, S53 and obtain equations for the unknowns $x$, $g$, $G_r(\omega_n)$, $\Sigma_r(\omega_n)$ and $\overline{\lambda}$. To do this, we need an expression for the inverse of a matrix in replica space. To obtain this, we use the following results for the product of 2 matrices in replica space (4). We consider the matrix $A_{ab}$ whose off-diagonal elements are parameterized by the function $a(u)$, $u \in [0,1]$, and the diagonal element $A_{aa} = \tilde{a}$. Similarly, we have matrices $B_{ab}$ and $C_{ab}$. Then if $C = AB$, we have

$$\begin{aligned} \tilde{c} &= \tilde{a}\tilde{b} - \langle ab \rangle \\ c(u) &= (\tilde{b} - \langle b \rangle)a(u) + (\tilde{a} - \langle a \rangle)b(u) - \int_0^u dv(a(u) - a(v))(b(u) - b(v)) \end{aligned} \quad [\text{S59}]$$

where

$$\langle a \rangle = \int_0^1 du\, a(u) \quad [\text{S60}]$$

So if $B = A^{-1}$ and $C = 1$, then $\tilde{c} = 1$ and $c(u) = 0$. Applying these results to the one-step replica symmetry breaking ansatz for $a(u)$ and $b(u)$, we obtain

$$\begin{aligned} b &= \frac{-a}{(a - \tilde{a})(a(1-x) - \tilde{a})} \\ \tilde{b} &= \frac{\tilde{a} + (x-2)a}{(a - \tilde{a})(a(1-x) - \tilde{a})} \end{aligned} \quad [\text{S61}]$$

where $a(u)$ and $b(u)$ have the same breakpoint $x$.

Applying Eqs. S61 to S51 we obtain the equations

$$\begin{aligned} G_r(i\omega_n) &= \frac{1}{i\omega + \overline{\lambda} - \Sigma_r(i\omega_n)}, \quad \text{for all } \omega_n \\ [gJG_r(i\omega_n = 0)]^2 &= 1 + \beta J^2 g^3 x G_r(i\omega_n = 0) \end{aligned} \quad [\text{S62}]$$

Notice that same equation applies for $G_r(i\omega_n)$ for $\omega_n = 0$ and $\omega_n \neq 0$, and this was the rationale for ansatz in Eq. S56.

It is useful to now introduce the dimensionless variable $\Theta$, defined by

$$\Theta = -JgG_r(i\omega_n = 0) \tag{S63}$$

Using Eq. S62, we see that $\Theta$ is related to $x$ by

$$\beta x = \frac{1}{Jg^2}\left(\frac{1}{\Theta} - \Theta\right), \tag{S64}$$

and so we eliminate $x$ in favor of $\Theta$. We also use the $\omega_n = 0$ equation in Eq. S62 to express $\overline{\lambda}$ in terms of $\Theta$ and $\Sigma_r(i\omega_n = 0)$.

Now the saddle-point equations are expressed in terms of $G_r(i\omega_n)$, $\Sigma_r(i\omega_n)$, $\Theta$, and $g$. The complete set of equations determining these parameters is

$$[G_r(i\omega_n)]^{-1} = i\omega_n - \frac{Jg}{\Theta} - [\Sigma_r(i\omega_n) - \Sigma_r(i\omega_n = 0)] \tag{S65}$$

$$\Sigma_r(\tau) = J^2\Big([G_r(\tau)]^2 G_r(-\tau) - 2gG_r(\tau)G_r(-\tau) - g[G_r(\tau)]^2$$
$$+ 2g^2 G_r(\tau) + g^2 G_r(-\tau)\Big) \tag{S66}$$

$$g - G_r(\tau = 0^-) = \kappa. \tag{S67}$$

These equations leave the parameter $\Theta$ undetermined.

Following Appendix B.2 in Ref. (2), we determine the value of $\Theta$ by the requirement that $G_r$ have a gapless spectrum. For a gapless spectrum, we expect the low frequency expansion for $G_r$ for real $\omega$

$$G_r(\omega + i\eta) = \frac{\Theta}{gJ} + (a + ib)\omega^\alpha \theta(\omega) + (a' + ib')|\omega|^\alpha \theta(-\omega) + \ldots \tag{S68}$$

where $\theta(\omega)$ is a step function, for some $\alpha > 0$, and with $b < 0$, $b' > 0$. Then from Eq. S66 we find that the leading singularity in $\Sigma_r$ is given by the last 2 terms in Eq. S66

$$\text{Im}\,\Sigma_r(\omega + i\eta) = g^2 J^2 \left[(2b - b')\omega^\alpha \theta(\omega) + (2b' - b)|\omega|^\alpha \theta(-\omega)\right] \tag{S69}$$

But from Eq. S65 we have

$$\text{Im}\,\Sigma_r(\omega + i\eta) = -\text{Im}\left[G_r(\omega + i\eta)\right]^{-1} = \frac{g^2 J^2}{\Theta^2}\left[b\,\omega^\alpha \theta(\omega) + b'\,|\omega|^\alpha \theta(-\omega)\right] \tag{S70}$$

Comparing Eqs. S69 and S70 we have

$$(1/\Theta^2 - 2)b + b' = 0$$
$$b + (1/\Theta^2 - 2)b' = 0 \tag{S71}$$

These equations are compatible only for $\Theta = 1, 1/\sqrt{3}$. The replica symmetric choice $\Theta = 1$ is excluded because it yields $b = b'$, which is not allowed by the positivity constraints on the spectral weight: $b$ and $b'$ need to have opposite signs. So the only choice is

$$\Theta = \frac{1}{\sqrt{3}}. \tag{S72}$$

## 4. Additional details for the metallic spin-glass phase

In this appendix we present details regarding the spectral densities in the metallic spin-glass phase. We show explicit form of the equations that are solved at zero temperature as well as details on spin and electron spectral densities at zero and finite temperatures.

**A. Details of the solution at zero temperature.** The analytic continuation of the equations Eq. (29)-Eq. (34) at zero temperature leads to the following equations for the spectral functions

$$\rho_r(\omega) = -\frac{1}{\pi}\frac{\Sigma_r''(\omega)}{(\omega - \frac{Jg}{\Theta} - (\Sigma_r'(\omega) - \Sigma_r'(0)))^2 + (\Sigma_r''(\omega))^2}; \tag{S73}$$

$$\rho_f(\omega) = -\frac{1}{\pi}\frac{\Sigma_f''(\omega)}{(\omega - \mu_f - (\Sigma_f'(\omega) - \Sigma_f'(0)))^2 + (\Sigma_f''(\omega))^2}. \tag{S74}$$

Maine Christos, Darshan G. Joshi, Subir Sachdev and Maria Tikhanovskaya

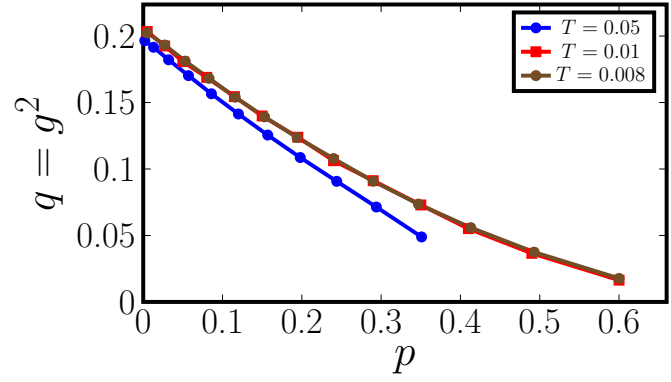

**Fig. S12.** Plot of spin-glass order parameter, $q = g^2$, as a function of doping at different temperatures.

Equations for self energies are more cumbersome and we list them in several parts. For the imaginary part of the boson self energy we have three terms $\Sigma_r''(\omega) = \Sigma_{r,1}''(\omega) + \Sigma_{r,2}''(\omega) + \Sigma_{r,3}''(\omega)$. The linear term reads

$$\Sigma_{r,1}''(\omega) = \pi J^2 g^2 [\rho_r(-\omega) - 2\rho_r(\omega)] \quad [\text{S75}]$$

$$\Sigma_{r,2}''(\omega) = \pi g \int_{-\infty}^{+\infty} d\omega_1 (\theta(-\omega_1)\theta(\omega_1 - \omega) - \theta(\omega_1)\theta(\omega - \omega_1)) \quad [\text{S76}]$$

$$\times \ (J^2 \rho_r(\omega_1)[-2\rho_r(\omega_1 - \omega) + \rho_r(\omega - \omega_1)] + kt^2 \rho_{\mathfrak{f}}(\omega_1)\rho_{\mathfrak{f}}(\omega_1 - \omega))$$

$$\Sigma_{r,3}''(\omega) = \pi \int_{-\infty}^{+\infty} d\omega_1 d\omega_2 (\theta(-\omega_1)\theta(-\omega_2)\theta(-\omega + \omega_1 + \omega_2) + \theta(\omega_1)\theta(\omega_2)\theta(-\omega_1 - \omega_2 + \omega))$$

$$\times \ (J^2 \rho_r(\omega_1)\rho_r(\omega_2)\rho_r(\omega_1 + \omega_2 - \omega) + kt^2 \rho_{\mathfrak{f}}(\omega_1)\rho_{\mathfrak{f}}(\omega_2)\rho_r(\omega_1 + \omega_2 - \omega))$$

Similarly, the imaginary part of the fermion self energy can be written as $\Sigma_{\mathfrak{f}}''(\omega) = \Sigma_{\mathfrak{f},1}''(\omega) + \Sigma_{\mathfrak{f},2}''(\omega) + \Sigma_{\mathfrak{f},3}''(\omega)$, in particular

$$\Sigma_{\mathfrak{f},1}''(\omega) = -\pi t^2 g^2 \rho_{\mathfrak{f}}(\omega) \quad [\text{S77}]$$

$$\Sigma_{\mathfrak{f},2}''(\omega) = \pi t^2 g \int_{-\infty}^{+\infty} d\omega_1 \rho_{\mathfrak{f}}(\omega_1) [\rho_r(\omega - \omega_1) - \rho_r(\omega_1 - \omega)](\theta(-\omega_1)\theta(-\omega + \omega_1) - \theta(\omega_1)\theta(-\omega_1 + \omega))$$

$$\Sigma_{\mathfrak{f},3}''(\omega) = \pi t^2 \int_{-\infty}^{+\infty} d\omega_1 d\omega_2 (\theta(-\omega_1)\theta(-\omega_2)\theta(-\omega + \omega_1 + \omega_2) + \theta(\omega_1)\theta(\omega_2)\theta(-\omega_1 - \omega_2 + \omega))$$

$$\times \ \rho_{\mathfrak{f}}(\omega_1)\rho_r(\omega_2)\rho_r(\omega_1 + \omega_2 - \omega)$$

We obtain the real parts of the self energies by using the Kramers-Kronig relation. The filling constraints read

$$g - \int_{-\infty}^{0} d\omega \rho_r(\omega) = \kappa - kp \quad [\text{S78}]$$

$$\int_{-\infty}^{0} d\omega \rho_{\mathfrak{f}}(\omega) = p \quad [\text{S79}]$$

We fix parameters to $J = t = 1$ and $\kappa = k = 1/2$ everywhere for the spin glass phase. The solution of these equations is showns in Fig. 8 for different $p$. For the boson spectral density $\rho_r(\omega)$ we see linear in frequency behavior at small frequencies.

**B. Spectral functions.** In this section we find spectral functions of gauge invariant observables: spin and electron spectral densities. We also compute the sum rules for these functions at finite and zero temperatures. Recall that in the spin-glass phase we use the $SU(M|M')$ representation where the holon is fermionic and spinons are bosonic operators. From Eq. 28,

$$G_c(\tau) = G_{\mathfrak{b}}(\tau) G_{\mathfrak{f}}(-\tau) = [G_r(\tau) - g] G_{\mathfrak{f}}(-\tau). \quad [\text{S80}]$$

Therefore the electron spectral density is

$$\rho_c(\omega) = -\frac{1}{\pi}\text{Im}[G_c(\omega + i0^+)] = g\rho_{\mathfrak{f}}(-\omega) + \int_{-\infty}^{\infty} d\omega_2 \ \rho_{\mathfrak{f}}(\omega_2)\rho_r(\omega + \omega_2)[n_B(\omega + \omega_2) + n_F(\omega_2)]. \quad [\text{S81}]$$

It is straightforward to obtain the sum rule,

$$\int_{-\infty}^{\infty} d\omega \ \rho_c(\omega) = \frac{1+p}{2}. \quad [\text{S82}]$$

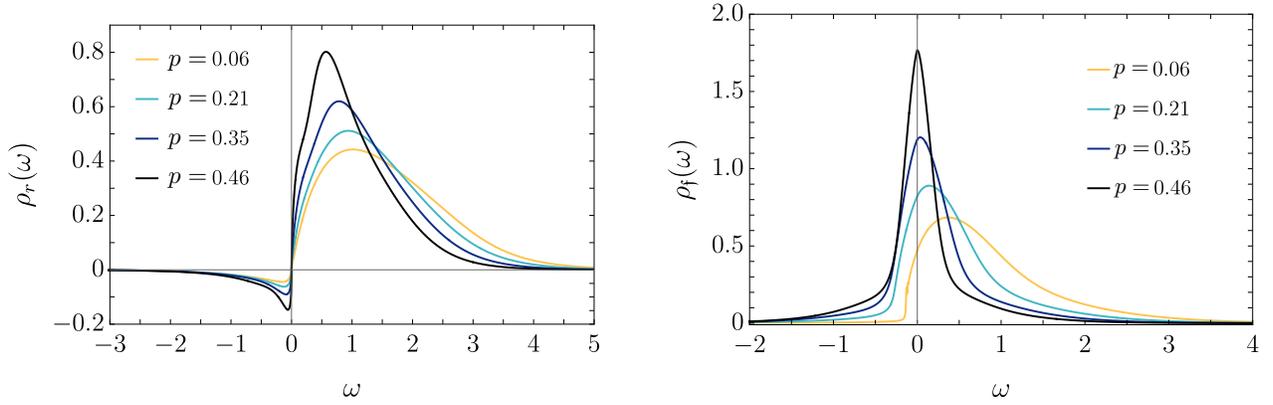

**Fig. S13.** Spectral functions obtained at zero temperature via real frequency analysis for several values of doping. Parameters are $J = t = 1$.

Recall that this is same as that obtained in the critical metal phase.

For the spin correlator we have,

$$\chi(\tau) = G_{\mathfrak{b}}(\tau)G_{\mathfrak{b}}(-\tau) = g^2 - g\left[G_r(\tau) + G_r(-\tau)\right] + G_r(\tau)G_r(-\tau)\,. \tag{S83}$$

Here, $\chi'' = \text{Im}[\chi(\tau)]$, where $\chi(\tau) = \langle \vec{S}(\tau) \cdot \vec{S}(0) \rangle$. This gives the spin spectral density,

$$\rho_s(\omega) = \frac{\chi''(\omega)}{\pi} = g^2 \beta \omega \delta(\omega) + g\left[\rho_r(\omega) - \rho_r(-\omega)\right] - \int_{-\infty}^{\infty} d\omega_2\ \rho_r(\omega_2)\rho_r(\omega+\omega_2)\left[n_B(\omega+\omega_2) - n_B(\omega_2)\right]\,. \tag{S84}$$

The sum rule for the spin spectral density at non-zero temperature is not easy to obtain and we do not have an exact expression for it. At zero temperature, however, we obtain

$$\int_0^\infty d\omega\ \rho_s(\omega) = \frac{(1-p)(3-p)}{4} - g^2\,. \tag{S85}$$

In addition, we have the usual sum rules,

$$\int_{-\infty}^{\infty} d\omega\ \rho_{\mathfrak{f}}(\omega) = \int_{-\infty}^{\infty} d\omega\ \rho_r(\omega) = 1\,. \tag{S86}$$

Solution of these equations for several values of doping is presented in Fig. S13.



\end{document}